\documentclass[11pt,letterpaper]{article}
\pdfoutput=1

\usepackage[section]{placeins}
\usepackage{jcappub}
\usepackage{natbib}
\usepackage{graphicx}
\usepackage{dcolumn}
\usepackage{bm}
\usepackage{epsfig}
\usepackage{amsmath}
\usepackage{color}
\usepackage{longtable}
\usepackage{epstopdf}
\usepackage{hyperref}

\newcommand\approxgt{\lower.6ex\hbox{$\sim$}\llap{\raise.4ex\hbox{$>$}}$\,$}
\newcommand\approxlt{\lower.6ex\hbox{$\sim$}\llap{\raise.4ex\hbox{$<$}}$\,$}
\newcommand{\be}{\begin{equation}}
\newcommand{\ee}{\end{equation}}
\newcommand{\bea}{\begin{eqnarray}}
\newcommand{\eea}{\end{eqnarray}}

\def\thetaB{\mbox{\boldmath$\hat\theta$}}
\newcommand\gsim{\mathrel{\rlap{\lower4pt\hbox{\hskip1pt$\sim$}}
        \raise1pt\hbox{$>$}}}
\newcommand\lsim{\mathrel{\rlap{\lower4pt\hbox{\hskip1pt$\sim$}}
        \raise1pt\hbox{$<$}}}
\def\thetaB{\mbox{\boldmath$\hat\theta$}}

\title{Sloan Digital Sky Survey III Photometric Quasar Clustering: Probing the Initial Conditions of the Universe}
\author[a,1]{Shirley Ho,\note{Corresponding author.}}
\author[a]{Nishant Agarwal,}
\author[b,c]{Adam D.\ Myers,}
\author[a]{Richard Lyons,}
\author[a]{Ashley Disbrow,}
\author[d]{Hee-Jong Seo,}
\author[e]{Ashley Ross,}
\author[f,g]{Christopher Hirata,} 
\author[h]{ Nikhil Padmanabhan,} 
\author[a]{Ross O'Connell,}
\author[f]{Eric Huff,}
\author[d]{David Schlegel,}
\author[i]{An\v{z}e Slosar,}
\author[f]{David Weinberg,}
\author[j]{Michael Strauss,}
\author[d,k]{Nicholas P.  Ross,}
\author[l,m]{Donald P. Schneider,}
\author[j]{Neta Bahcall,}
\author[n]{J. Brinkmann,}
\author[o]{Nathalie Palanque-Delabrouille,}
\author[o]{and Christophe Y\`{e}che} 

\affiliation[a]{McWilliams Center for Cosmology, Department of Physics, Carnegie Mellon University, 5000 Forbes Avenue, Pittsburgh, PA 15213, USA}
\affiliation[b]{Department of Physics and Astronomy, University of Wyoming, Laramie, WY 82071, USA}
\affiliation[c]{Max-Planck-Institut f\"ur Astronomie, K\"onigstuhl 17, D-69117, Heidelberg, Germany}
\affiliation[d]{Lawrence Berkeley National Laboratory, 1 Cyclotron Rd, Berkeley, CA 94702}
\affiliation[e]{Institute of Cosmology \& Gravitation, Dennis Sciama Building, University of Portsmouth, Portsmouth, PO1 3FX, UK}
\affiliation[f]{Department of Astronomy, Ohio State University, 140 West 18th Avenue, Columbus, OH 43210, USA}
\affiliation[g]{Department of Physics, Ohio State University, 140 West 18th Avenue, Columbus, OH 43210, USA}
\affiliation[h]{Department of Physics and Astronomy, Yale University, New Haven,  CT 06520}
\affiliation[i]{Brookhaven National Laboratory, Bldg. 510, Upton NY 11375, USA}
\affiliation[j]{Department of Astrophysical Sciences, Princeton University, Princeton, NJ 08544}
\affiliation[k]{Department of Physics, Drexel University, 3141 Chestnut Street, Philadelphia, PA 19104, USA }
\affiliation[l]{ Department of Astronomy and Astrophysics, The Pennsylvania State University, University Park, PA 16802}
\affiliation[m]{Institute for Gravitation and the Cosmos, The Pennsylvania State University, University Park, PA 16802, USA}
\affiliation[n]{Apache Point Observatory,  P.O. Box 59, Sunspot, NM 88349-0059, USA} 
\affiliation[o]{CEA, Centre de Saclay, Irfu/SPP, F-91191 Gif-sur-Yvette, France} 

\emailAdd{shirleyh@andrew.cmu.edu}

\begin{document} 

\maketitle

\newpage

\abstract{Abstract} 

The Sloan Digital Sky Survey has surveyed 14,555 square degrees of the sky, and delivered over a trillion pixels of imaging data. We present the large-scale clustering of 1.6 million quasars between $z=0.5$ and $z=2.5$ that have been classified from this imaging, representing the highest density of quasars ever studied for clustering measurements. This data set spans $\sim$ 11,000 square degrees and probes a volume of  $80 h^{-3}$ Gpc$^3$. In principle, such a large volume and medium density of tracers should facilitate high-precision cosmological constraints. We measure the angular clustering of photometrically classified quasars using an optimal quadratic estimator in four redshift slices with an accuracy of $\sim 25\%$ over a bin width of $\delta_l \sim 10-15$ on scales corresponding to matter-radiation equality and larger ($\ell \sim 2-30$).

Observational systematics can strongly bias clustering measurements on large scales, which can mimic cosmologically relevant signals such as deviations from Gaussianity in the spectrum of primordial perturbations. 
We account for systematics by employing a new method recently proposed by Agarwal et al.\ (2014) to the clustering of photometrically classified quasars. We carefully apply our methodology to mitigate known observational systematics and further remove angular bins that are contaminated by {\em unknown} systematics. Combining quasar data with the photometric luminous red galaxy (LRG) sample of Ross et al.\ (2011) and Ho et al.\ (2012), 
and marginalizing over all bias and shot noise-like parameters, we obtain a constraint on local primordial non-Gaussianity of $f_{\rm NL} = -113^{+154}_{-154} \ (1\sigma {\rm \ error})$. We next assume that the bias of quasar and galaxy distributions can be obtained independently from quasar/galaxy-CMB lensing cross-correlation measurements 
(such as those in Sherwin et al.\ (2013)). 
This can be facilitated by spectroscopic observations of the sources, enabling the redshift distribution to be completely determined, and allowing precise estimates of the bias parameters. 
In this paper, if the bias and shot noise parameters are fixed to their known values (which we model by fixing them to their best-fit Gaussian values), we find that the error bar reduces to $1\sigma \simeq 65$. 
We expect this error bar to reduce further by at least another factor of five if the data is free of any observational systematics. We therefore emphasize that in order to make best use of large scale structure data we need an accurate modeling of known systematics, a method to mitigate unknown systematics, and additionally independent theoretical models or observations to probe the bias of dark matter halos.


\section{Introduction}
\label{sec:intro}

Maps of the distribution of light have long been used to probe the structure of the Universe. In 1917, Einstein wrote of the distribution of stars as possibly being uniform when averaged over large distances \citep{peebles73}. In a similar vein, in 1926 Hubble famously measured the distribution of faint nebulae to test the uniformity of the Cosmos \citep{peebles73}. As the structure of the Universe over increasingly large volumes has become better understood, the distribution of light from objects such as galaxies has remained a powerful cosmological probe \citep{peebles73,groth73,wang99,hu99,eisenstein99}.

Quasars classified in wide-area imaging surveys are  obvious tracers with which to probe the distribution of light across even larger volumes \citep{myers06,myers07a,myers07b,Slosar:2008hx}. Smoothed over {\em sufficiently}  large scales, we expect the number density of quasars to have a simple relationship to the underlying matter density. This implies that quasar clustering on large scales is directly related to the clustering of the dark matter in which quasars are embedded. Quasar clustering is thus a sensitive probe of the structure and evolution of the Universe.

Hidden in the ever-increasing volumes encompassed by large imaging surveys is a wealth of cosmological information that has yet to be fully exploited. Because gravitational and astrophysical effects influence the evolution of clustering on ``small-to-moderate'' scales, the very-large-scale clustering of any mass tracer --- usually as characterized by its power spectrum --- can constrain the {\em primordial} potential of the Universe \citep[e.g.,][]{dalal08}. Deviations from a Gaussian distribution of this potential are typically measured using various $f_{\rm NL}$ parameters for different shapes of the three-point function, which parameterize the contribution of non-Gaussian modes to the primordial potential field \citep{komatsu02,Maldacena:2002vr}. This quantity is a powerful cosmological observable, as different inflationary scenarios produce primordial potentials that deviate from Gaussianity at different levels. 

In the so-called squeezed limit of the three-point function it is found that significant non-Gaussanity directly affects the abundance and clustering of virialized objects \cite{Grinstein:1986en,Matarrese:1986et,dalal08,Matarrese:2008nc,Slosar:2008hx}. A very useful example of a non-Gaussian scenario that affects the clustering of dark matter halos is the ``local'' ansatz, in which the gravitational potential is a simple non-linear function of the local value of a Gaussian field. The predicted form of the halo power spectrum in the presence of primordial non-Gaussianity has been used to constrain local $f_{\rm NL}$ using data from different tracers of large scale structure, e.g., see \cite{Slosar:2008hx, Xia:2010yu,Xia:2010pe,Xia:2011hj,Ross:2012sx,Giannantonio:2013uqa,Karagiannis:2013xea,Leistedt:2014zqa}. The local $f_{\rm NL}$ parameter allows us to directly probe different models of inflation. A small value of $f_{\rm NL} \lesssim 1$ can be explained using standard, slow-roll, single-field inflation \cite{Maldacena:2002vr,Creminelli:2004yq,Creminelli:2011rh,Pajer:2013ana}. Higher values, however, point towards a more general model of inflation, such as multi-field inflation \cite{Linde:1996gt,Bernardeau:2002jy,Seery:2005gb,Rigopoulos:2005us,Vernizzi:2006ve,Battefeld:2006sz,
Yokoyama:2007uu,Yokoyama:2007dw,Sasaki:2008uc,Byrnes:2008wi,Byrnes:2010em,Kim:2010ud,
Peterson:2010mv,Elliston:2011et, Dias:2012nf}, the curvaton scenario \cite{Lyth:2002my,Ichikawa:2008iq,Beltran:2008aa,Chambers:2009ki,Byrnes:2009pe,Enqvist:2009ww,
Alabidi:2010ba}, or a single field model with a modified initial state \cite{Chen:2006nt,Holman:2007na,Agullo:2010ws,Ashoorioon:2010xg,Ganc:2011dy,Kundu:2011sg,
Chialva:2011hc,Agarwal:2012mq,Flauger:2013hra,Aravind:2013lra,Ashoorioon:2013eia}. 

The Planck satellite has recently placed the most stringent constraints on primordial non-Gaussianity using measurements of temperature anisotropies in the cosmic microwave background (CMB) \cite{Ade:2013ydc}. The constraints, however, have not yet excluded any complete class of inflationary models. It is expected that large scale structure surveys in the future may be able to place even stronger constraints on primordial non-Gaussianity \cite{Fedeli:2009fj,Carbone:2010sb,Cunha:2010zz,Fedeli:2010ud,Giannantonio:2011ya, Pillepich:2011zz,Huang:2012mr,Carrasco:2012cv}. In this paper we use recent photometric data of quasars in the Sloan Digital Sky Survey Data Release Eight (SDSS DR8) sample to constrain background cosmology and primordial non-Gaussianity.

SDSS-III DR8 \citep{aihara11,eisenstein11} has imaged $14,555$\,$\rm{deg}^2$ of the sky in five bands (ugriz). The precision, depth, and wavelength coverage of SDSS imaging, which is unparalleled by any similarly large digital sky survey, allows the construction of  a large, uniform sample of photometrically classified quasars, with photometric redshifts, to faint fluxes, as described in \S\ref{sec:qsodef}. Spectroscopy of a uniform subsample of 1\% of these photometrically classified quasars \citep{pal12} allows us to carefully  characterize and calibrate our sample. 

In this paper, we make use of DR8 photometrically classified quasars to derive the most accurate and precise measurement of the quasar angular power spectrum achieved to date. Due to non-linear evolution, the quasar density field is not Gaussian on small scales; however, at the large scales we consider, the field should be close to Gaussian. We therefore measure the angular power spectrum of quasars using an optimal quadratic estimator, which provides the maximum available information for a Gaussian field. With a large volume such as that covered by the SDSS, the effects of large scale systematics are non-negligible \cite{ross11,ho12}. The multi-epoch imaging available in DR8 allows the investigation of how variations in (e.g.) seeing and sky  brightness can mimic non-linear clustering on large scales. We employ an innovative method that cross-correlates maps of systematics with maps of the quasar density to gauge and correct our large scale clustering measurements for non-cosmological contributions \cite{ross11,ho12}. In addition, we use cross-correlations between different quasar redshift slices to define cuts on the angular power spectrum due to unknown systematics \cite{agarwal13}.

The paper is organized as follows: \S\ref{sec:sample} describes the construction of the sample of photometrically classified quasars; \S\ref{sec:angular} presents the theory and measurement of the angular power spectrum; \S\ref{sec:sys} discusses various potential systematics; \S\ref{sec:theory_sys} describes the method we apply to mitigate observational systematics and characterize unknown systematics and  \S\ref{sec:results} summarizes the cosmological constraints themselves. We conclude in \S\ref{sec:discuss}. 


\section{The Data }
\label{sec:sample}

\subsection{SDSS Observations}

During its imaging phase, the SDSS \citep[][]{york00} mapped over a quarter of the sky using the dedicated Sloan Foundation 2.5-meter telescope located at Apache Point Observatory in New  Mexico \citep{gunn06}. A drift-scanning mosaic CCD camera \citep{gunn98,gunn06} imaged the sky in five photometric band-passes \citep{fukugita96,smith02} to a limiting magnitude of $r\sim 22.5$. The imaging data was processed through a series of pipelines that perform astrometric calibration \citep{pier03}, photometric reduction \citep{lupton02}, and photometric calibration \citep{padmanabhan08}. In particular, the third incarnation of the SDSS includes the Baryon Oscillation Spectroscopic Survey \citep[SDSS-III/BOSS;][]{aihara11,eisenstein11}, which added $3000$\,$\rm{deg}^2$ of new imaging to SDSS-I/II.

The availability of a large, uniform set of imaging data makes BOSS an obvious resource for efficiently photometrically classifying significant numbers of quasars \citep[as in][]{Ric04,Ric09}. In addition, SDSS-III is now following up selections of targets selected from SDSS imaging, which are designated for spectroscopy 
using an adaptive tiling algorithm based on \cite{blanton03}, and observed with a pair of fiber-fed spectrographs \citep{bolton12, smee13}.  This spectroscopy is vital to better characterizing and calibrating large maps of photometrically classified sources. A summary of the survey design of BOSS appears in \cite{eisenstein11}, and a full description is provided in \cite{dawson13}.

\subsection{Photometrically Classified Quasars from DR8 Imaging}
\label{sec:qsodef}

We use a photometric quasar catalog constructed from SDSS DR8 imaging using {\it Extreme Deconvolution} (XD).\footnote{XD \protect \citep{xd11} rapidly and robustly models the density distribution of a parameter (e.g., astronomical sources in color space) as a sum of Gaussians convolved with measurement errors.} This catalog was created by applying XD to the 103,601 $z > 0.3$ spectroscopically confirmed quasars in the SDSS DR7 quasar catalog \citep{Sch10} to model the density distribution of quasars in ($ugriz$) flux-redshift space. This density is compared to a model of the density of non-quasars in flux-space drawn from point sources in $150\,{\rm deg}^2$ of SDSS Stripe 82 that do not substantially vary in flux \citep[see][for more details]{bovy11,Kir11}. By applying these model densities and integrating the quasar flux-redshift 
density over different redshift ranges, XD can be used to calculate the probability that any source drawn from SDSS $ugriz$ imaging is a quasar in a given redshift range. This process, which is referred to as ``XDQSOz'', is described in detail in \citep{bovy12}. We will refer to the probability of being a quasar as calculated by XDQSOz over all possible redshifts ($0 < z < \infty$) as the ``XDQSOz probability.''

A similar approach to the XDQSOz technique that was used to construct the catalog we use in this paper  
was used to derive probabilities for targeting for the BOSS CORE quasar sample \citep[e.g.][]{ross12,white12}. For instance, our sample employs the same flag cuts as used for the CORE BOSS quasar sample to remove imaging glitches, including the imposed (dereddened) Point Spread Function (PSF) magnitude (for more description on magnitude systems in SDSS, please refer to \cite{lupton02}) limits of {\tt ($g \leq 22~||~r \leq 21.85)~\&\&~i \geq 17.8$}.\footnote{These flag cuts are equivalent to {\tt good==0} described in Appendix A of \citep{bovy11}.} There are some notable differences in our sample. The catalog we use is constructed using the XDQSOz formalism of \citep{bovy12}; rather than XDQSO \citep{bovy11}. The catalog we use is not restricted to objects with an XDQSO probability of ${\tt pqsomidz > 0.424}$ in the ``mid-$z$" redshift range ($2.2 \leq z \leq 3.5$). Rather,  we use sources with an XDQSOz quasar probability of $> 0.9$ across all redshifts. Finally, we impose an additional magnitude cut of $i < 21.5$ to guard against faint, unresolved galaxies contaminating our sample. At $i = 21.5$, fewer than 4\% of imaged SDSS sources with an XDQSOz quasar probability of $> 0.9$ are point-like galaxies \citep[see Fig.\ 12 of][]{bovy12}.

After applying flag cuts and magnitude limits, the XDQSOz-constructed catalog we consider contains 421{,}121 (1{,}615{,}226) photometrically classified objects with a quasar probability threshold of  0.9 (0.5). XDQSOz can integrate probabilities over any redshift range, providing full photometric redshift PDFs (Probability Distribution Functions) for each of our photometrically classified quasars. We frequently use the most probable, or ``peak'' redshift of these PDFs to represent a single ``photometric redshift".


\subsection{Deep Spectroscopy of a Complete Quasar Sample}

BOSS collaborators can submit ancillary programs, which utilize fibers that would otherwise be unassigned to targets \citep{dawson13}. One such program surveyed a high-completeness, high-fiber-density sample of quasars across a large redshift range to fainter limits than the main BOSS quasar samples \cite{pal12}. This program was augmented by observations to yet fainter limits on the 6.5-meter MMT \citep[e.g.][]{west97} using Hectospec \citep[e.g.][]{Fab05}, as well as by standard quasar targets color-selected for BOSS \citep{pal11}. 

The BOSS ancillary program effectively reached a limit of $g\sim22.5$.\footnote{No actual limit beyond the SDSS imaging depth was imposed, but standard BOSS spectra have insufficient signal-to-noise to identify quasars beyond $g\sim22.5$.} The relevant observations are on BOSS plates numbered 5141 through 5147 and correspond to the  chunk ``boss21''.\footnote{The definition of a chunk is discussed in more detail in \cite{ross12}.} The targeting sample for this BOSS ancillary program consisted of all point sources in the SDSS stripe 82 coadd \citep{annis11}. This sample was then culled to objects that exhibit significant variability \citep[$y_{\rm NN} > 0.5$; where $y_{\rm NN}$\footnote{$y_{\rm NN}$  is basically a probability output by  the Neural Network trained and tested using variability data in \protect \cite{pal11}.}  is as defined in][]{pal11} and meet the color criterion $c_3 < (1 - (c_1/ 3))$ where $c_1$ and $c_3$\footnote{$c_1$ and $c_3$ are linear functions of colors (u-g), (g-r), and (r-i) of the object.} are as defined in \cite[][their Eq.\ (13)]{fan99}.

The MMT program was limited to $g < 23$ with the same color and variability cuts {\em for point sources} as imposed in the BOSS ancillary program. But, the MMT program also incorporated {\em resolved} sources from 
the SDSS Stripe 82 coadd. Extended sources were targeted if they met $y_{\rm NN} > 0.8$ and $c_3 < (0.6 - (c_1/ 3))$, with $y_{\rm NN}, c_1$, and $c_3$ as defined in the previous paragraph.


These programs, and the resulting highly complete sample of 1877 spectroscopically confirmed quasars, are described in full in \cite{pal12}. We use this sample of confirmed quasars to better characterize the redshift  distribution of our photometrically-classified quasar catalog.

\subsection{Angular and Redshift Distributions}

To interpret the clustering of any sample, one must characterize the expected distribution of the sample as if it is completely random. This involves understanding both the angular and radial selection function in addition to the expected quasar density, which is characterized by its mean density.

\begin{figure}
\begin{center}
\includegraphics[width=3.0in]{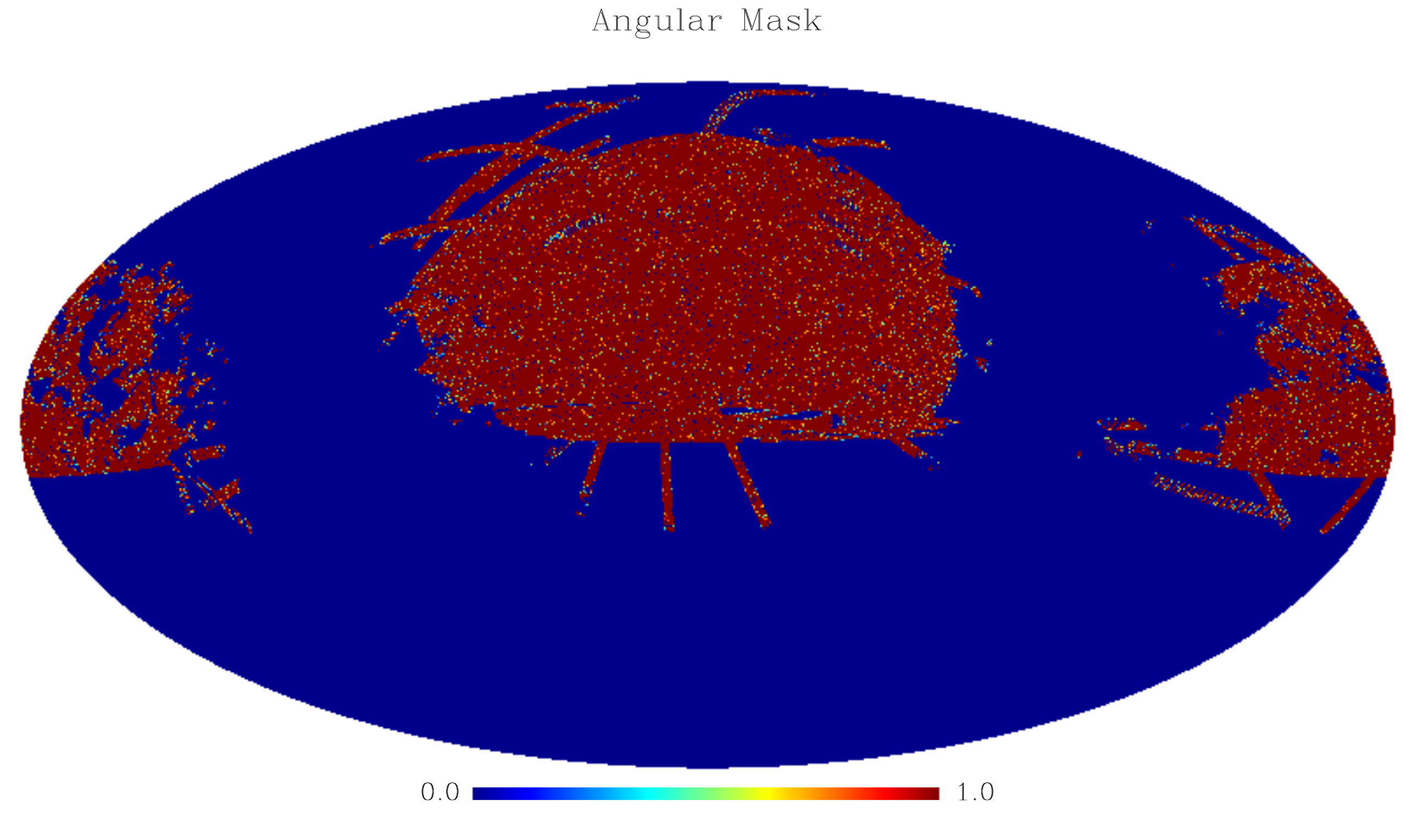}
\end{center}
\caption{The preliminary imaging mask after applying primary selection cuts, such as cuts on seeing and the 
bright star mask, to the full imaging mask.}
\label{fig:qsomask}
\end{figure}

To characterize the angular window function, we generate the complete angular mask of the survey following the procedures described in \cite{ho12} and \cite{aihara11}. To create a more restrictive mask which is catered towards photometric quasars, we exclude regions where the SDSS imaging quality indicator SCORE\footnote{Please refer to \cite{aihara11} for the specific definition of SCORE.} $> 0.5$, $E(B-V)> 0.08$  \citep{scranton02,myers06,ross06,padmanabhan07,ho08},  where seeing in the $i$-band exceeds $2.0 ''$ (FWHM), and regions around stars in the Tycho astrometric catalog \citep{hog00}. The final angular selection function covers a solid angle of $\sim 11,000$ square degrees, and  is shown in Fig.\ \ref{fig:qsomask}. 

\begin{figure}
\begin{center}
\includegraphics[angle=0, width=4.0in]{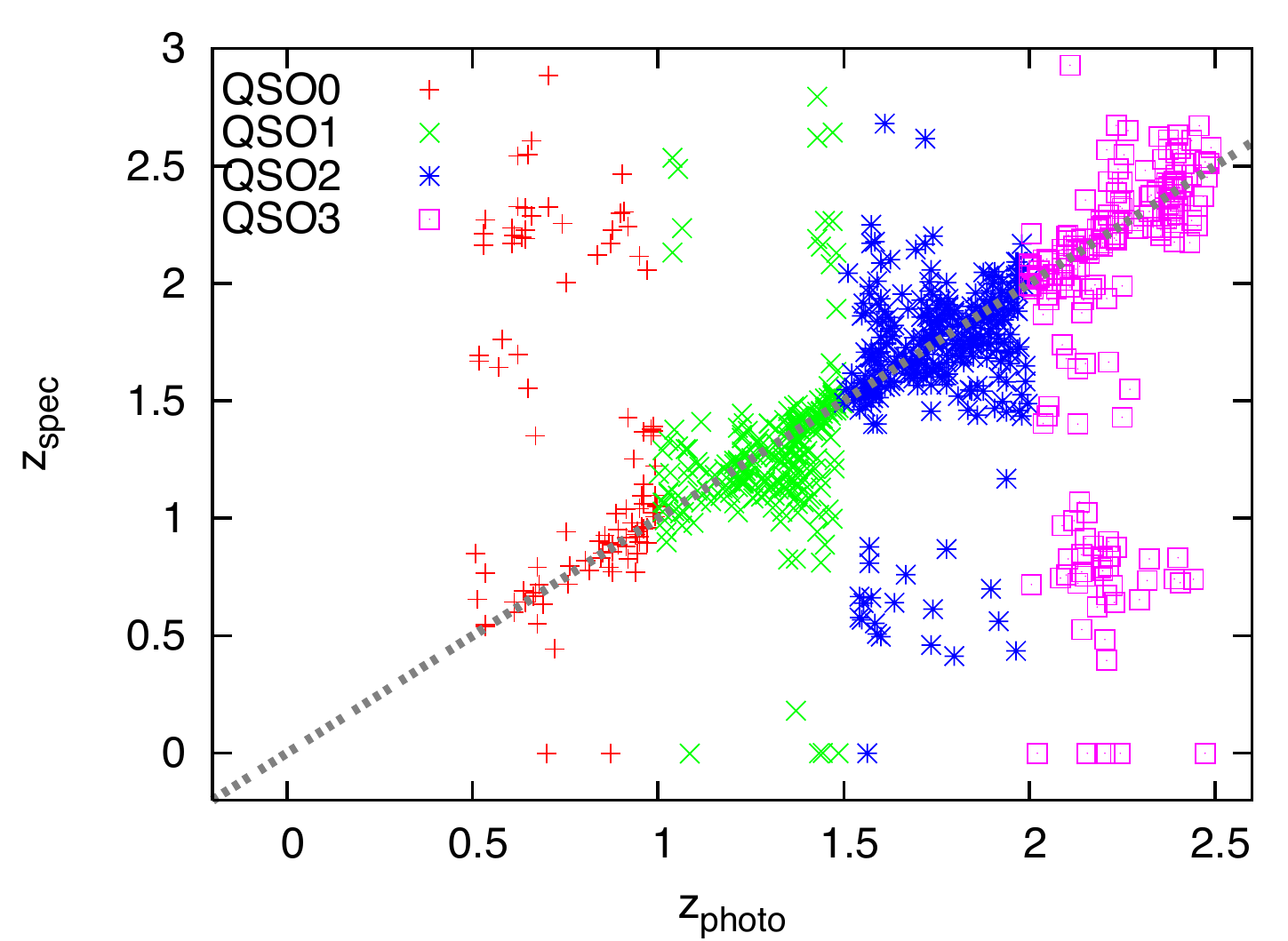}
\end{center}
\caption{The photometric vs. spectroscopic redshift distribution of SDSS-III photometric quasars that had been observed spectroscopically as described in \S\ref{sec:qsodef}. The four redshift slices are designated by QSO0-3. One can see that the dispersion for QSO1 and QSO2 is significantly smaller than both QSO0 and QSO3.  }
\label{fig:zslicecmass}
\end{figure}

For every object, the photometric redshift and XDQSOz probability of being a quasar were determined as described in \S\ref{sec:qsodef}.
As seen in Fig.\ \ref{fig:zslicecmass}, there are significant numbers of outliers in the relationship between photometric and spectroscopic redshift (often called ``catastrophic failures''). This  feature is typical of quasar photometric redshifts \citep[e.g.][]{bud01,ric01,Ric04,wei04,bal08,Ric09} and describing the scatter including the outliers would not be very meaningful. Therefore, we use a ``sigma-clip cut'' where we exclude objects more than 1-sigma from the mean and then recalculate the scatter. We calculate $\sigma(z_{\rm spec}-z_{\rm photo})$ for each redshift slice after the sigma-clip cut and document the catastrophic outlier rate ($R_{\rm fail}$) in Table\ \ref{tab:qsodndz}.



\begin{table}[!h]
\begin{center}
\begin{tabular}{|c|c|c|c|c|c|}
\hline
Sample & $z_{\rm mid}$ & $N_{\rm spec} ({\rm before~cut})$  & $\sigma(z_{\rm spec}-z_{\rm photo}) $ & $R_{\rm fail}$ \\
\hline
QSO0 & 0.75   & 81 (105)   &  0.406  &  22\% \\
QSO1 & 1.25   & 223 (247)  &  0.098  & 9.3\% \\
QSO2 & 1.75   & 251 (282)  &  0.117  &  10.7\% \\
QSO3 & 2.25   & 146 (188)  &  0.167  &  22\% \\
\hline
\end{tabular}
\caption{A rough characterization of the photometric redshift accuracy for our photometrically classified quasars based on a nearly complete sample of spectroscopically confirmed quasars. We choose to remove catastrophic failures, which are more than 1-sigma from the mean, and then recalculate the scatter after the removal.}
\label{tab:qsodndz}
\end{center}
\end{table}


We create pixelized maps of the quasar number overdensity (weighted by the XDQSOz probability of being a quasar), $\delta_g=\delta n/\bar n$, using a HEALPix pixelization \citep{gorski99} of the sphere. With 12{,}582{,}912 pixels over the whole sphere (HEALPix resolution 10, nside=1024), each pixel covers a solid angle of 11.8 $\rm{arcmin}^2$. These pixelized maps are used directly to compute the angular power spectra using an optimal quadratic estimator. The optimal quadratic estimator does not down-sample input pixelized maps; it computes the covariance matrix directly from these pixelized maps, as discussed further in \S\ref{sec:est_theory}.

\begin{figure}
\begin{center}
\includegraphics[width=3.0in]{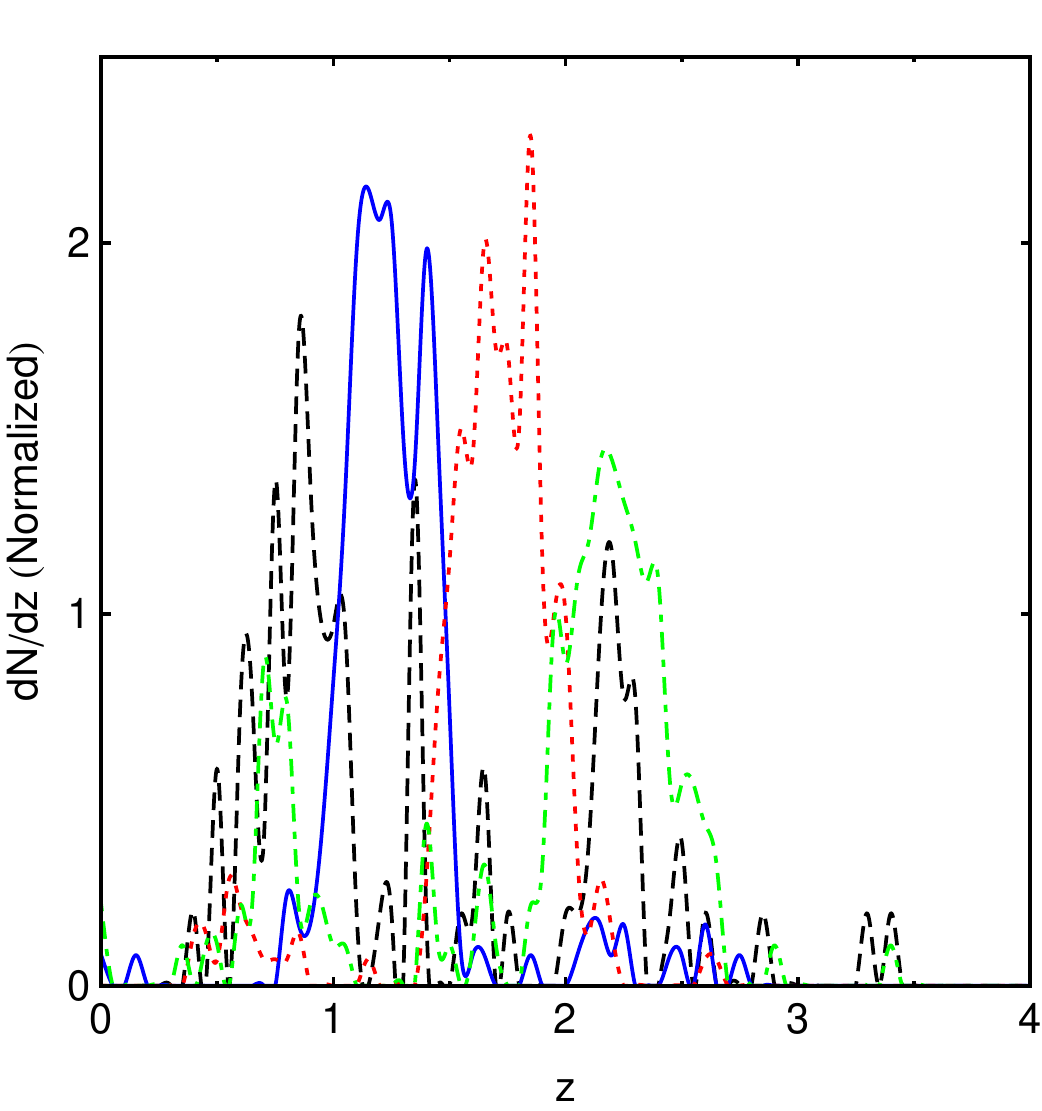}
\end{center}
\caption{The redshift distribution of the photometric quasar sample when we match the objects with an unbiased subsample as described in \S\ref{sec:qsodef} from SDSS-III BOSS. The four different colors designate the four different redshift samples. We can see that there is quite a bit of dispersion, especially in the lowest redshift slice (QSO0) and the highest redshift slice (QSO3).}
\label{fig:dndz}
\end{figure}

The quasar sample is divided into four photometric redshift slices of thickness $\Delta z = 0.5$ starting at $z=0.5$ and ending at $z=2.5$. We designate these samples QSO0 through QSO3 (see Table\ \ref{tab:qsodata} for details). The underlying redshift distributions for each slice are calculated using spectroscopically confirmed quasars from the (g~\approxlt\,22.5) BOSS ancillary program, as described in \S\ref{sec:qsodef}. There is a small portion (2.6\%) of photometric quasars that do not fall into this redshift range and they are not  included in this analysis. 

The redshift distribution of the sample is plotted in Fig.\ \ref{fig:dndz}. We can see that although the majority of objects in any photometric redshift bin is in its corresponding true redshift bin, a significant fraction of
objects falls into neighboring bins. The comparisons of these photometric redshifts to the spectroscopic redshifts (obtained via SDSS III spectra) are plotted in Fig.\ \ref{fig:zslicecmass}, while properties of the different slices are summarized in Table\ \ref{tab:qsodata}. 

\begin{table}[!h]
\begin{center}
\begin{tabular}{|c|c|c|c|c|c|}
\hline
Label & $z_{\rm mid}$ & $N_{\rm gal}$ & $l_{\rm max}$ & $b_{1}$ & $b_{1}$ \\
& & & & (initial) & (final) \\
\hline
QSO0 & 0.75 & 47710 & 189 & $3.01^{+0.18}_{-0.15}$ & $2.19^{+0.47}_{-2.09}$ \\
& & & & & Median: 2.57 \\
QSO1 & 1.25 & 142096  & 278 & $2.22^{+0.11}_{-0.11}$ & $2.06^{+0.08}_{-0.08}$ \\
QSO2 & 1.75 & 148166  & 346 & $2.45^{+0.14}_{-0.14}$ & $2.32^{+0.11}_{-0.09}$ \\
QSO3 & 2.25 & 71942  & 400 & $3.64^{+0.31}_{-0.31}$ & $3.37^{+0.20}_{-0.18}$ \\
\hline
\end{tabular}
\caption{Properties of the four $\Delta z=0.5$ redshift slices; $z_{\rm mid}$ is the midpoint of the redshift interval. Bias parameters (with $1\sigma$ errors) are deduced by marginalizing over all other cosmological parameters, combining WMAP9 + SN + DR8 (QSO) data sets. In the fifth column we use all available $\ell$ bins in the range $30 \leq \ell \leq \ell_{\rm max}$ and in the last column we report the final bias from using multipoles in the range $10 \leq \ell \leq \ell_{\rm max}$ that are not significantly contaminated with unknown systematics (see \S\ref{sec:theory_sys}).}
\label{tab:qsodata}
\end{center}
\end{table}

\subsection{Sample Systematics}

There are a number of potential systematic effects in photometric samples that contaminate clustering: stellar contamination and obscuration, seeing variations, sky brightness variations, extinction, and color offsets (such as those described in \cite{schlafly10}). These potential systematics were discussed extensively in \cite{ho12}, so we will concentrate on the various systematics that affect angular power spectra in the range of interest of our science analysis. 


\section{The Angular Power Spectrum}
\label{sec:angular}

As was noted in \S\ref{sec:intro}, the angular power spectrum contains information from both the growth and the expansion of the Universe, encoded in two standard rulers --- baryon acoustic oscillations and the matter radiation equality turn-over scale (i.e. the shape of the power spectrum). In this section we will briefly summarize both the theory and the computation of angular power spectra \citep[see, e.g., ][for a more detailed description]{ho12}. 

\subsection{From Quasar Distributions to the Angular Power Spectrum}
\label{sec:theory1}

The intrinsic angular quasar fluctuations are given by
\begin{equation}
	q(\thetaB)= \int {\rm d}z \,b(z) N(z) \delta(\chi(z)\thetaB,z) \, ,
\label{dg}
\end{equation}
where $b(z)$ is a linear bias factor, which is assumed to be scale-independent, relating the quasar overdensity to the mass overdensity, i.e.,  $\delta_q =b\delta$, $N(z)$ is the normalized selection function, and $\chi(z)$ is the comoving distance to redshift $z$. We focus on the auto power spectrum of quasars,
\begin{equation}
\label{cc1}
	C^{qq}(\ell)=\frac{2}{\pi} \int {\rm d}k \, k^2 P(k) [q]_\ell (k) [q]_\ell(k) \, ,
\end{equation}
where $P(k) = P(k,z=0)$ is the matter power spectrum today as a function of the wave number $k$, and the function $[q]_\ell$ is
\begin{equation}
\label{cc2}
	\left[q\right]_\ell(k)=\int {\rm d}z \, b_i(z) N(z) D(z)j_\ell(k\chi(z)) \, ,
\end{equation}
and $j_{l}(x)$ is the $l^{\rm th}$ order spherical Bessel function.  

For an auto-correlation, applying the Limber approximation \citep{loverde08} changes Eq.\ (\ref{cc1}) to
\begin{equation} 
	C^{qq}_\ell=\int {\rm d} z \frac{1}{\chi^2(z)} b^2(z) N^2(z) P(k,z) \, .
\label{cc2}
\end{equation}
Note that the Limber approximation introduces an error on the order of $1/\ell^2$ \citep{loverde08}, therefore, for $\ell $$<$ 10, we will face an ${\cal O}(1\%)$ error due to the Limber approximation. In this paper we only use $\ell$ $\geq$ 10 in all cases. 



For a cross-correlation between two different large scale structure samples (be it different selection functions, redshift distributions, biases etc.), we can write the cross-correlation as
\begin{equation} 
	C^{qq'}_\ell= \int {\rm d}z \frac{1}{\chi^2(z)} b(z) b'(z) N (z) N^{'}(z)  P(k,z) \, ,
\end{equation}
where $q'$ can have different biases, redshift dependence, etc. 

We have not yet distinguished between the quasar and matter angular power spectrum. We simply assume that
\begin{equation}
\label{eq:non-linear_eq}
	C_{q}(\ell) = b_{q}^{2} C_{\ell} + N_{\rm{shot}} + a \, ,
\end{equation}
where $C_{q}(\ell)$ and $C_{\ell}$ are the quasar and matter angular power spectra; $b_{q}$ is the linear quasar bias, $N_{\rm{shot}}$ is a constant shot noise term which is estimated by the optimal quadratic estimator, and $a$ is a constant shot noise-like term that is usually added to obtain a better fit to the non-linear power spectrum \cite{dePutter:2012sh}. This is a good approximation on large scales, but breaks down on smaller scales \citep{crocce06, crocce08, sanchez08,sanchez09,parfrey10,ho12, paranjape12, musso12}. Throughout the paper, we adopt this linear, redshift-independent (within our redshift slice) bias model. We set the non-linear fitting parameter $a$ to zero due to large error bars in the data. The bias for each redshift slice is fit as an extra parameter in Cosmological Monte Carlo (COSMOMC; \cite{lewis02}) chains to ensure that we do not systematically prejudice our cosmological models by fixing a pre-computed bias. We do not include the effects of magnification bias in this analysis as the paper aims to constrain the power in large scale through the auto-correlation of the quasar over density field. Cosmic magnification affects mostly the mid to small scales at the auto-correlation, although the effects are more significant in cross-correlations \citep{Leistedt:2013gfa}. 
 
\subsection{Redshift-Space Distortions}
\label{sec:theory2}

As described briefly in \cite{ho12}, we have investigated  the effects of redshift space distortions (RSDs) and found that non-linearities related to RSDs are not relevant on the scales with which we are concerned. We  only include the linear RSD effect, following \cite{padmanabhan07}. 

To be complete, let us review some of the important details from \cite{padmanabhan07}:
\be
	1+q(\thetaB) = \int \, {\rm d}\chi\,N(s)\,[1+\delta(\chi\thetaB,\chi)] \, ,
\label{eq:deltag_red}
\ee
where we have now written the normalized selection function as a function of redshift-space distance, $s = \chi + {\bf v}\cdot\thetaB$ with the peculiar velocity component, ${\bf v}$. Assuming the peculiar velocities are small compared with the thickness of the redshift slice, we Taylor expand the selection function to linear order,
\be
	N(s) \approx  N(\chi) + \frac{{\rm d}N}{{\rm d}\chi}({\mathbf v}\cdot\thetaB) \, .
\ee
Substituting this expression into Eq.\ (\ref{eq:deltag_red}), we express the 2D quasar density field as two separate terms, $q = q^{0} + q^{r}$, where $q^{0}$ is the term discussed in the previous section, while $q^{r}$ is the linear RSD correction. We can then use the linear continuity equation to derive the Legendre coefficient as
\be
	\delta_{q}^{r}(\ell) = i^{\ell}\,\int \frac{{\rm d}^{3}k}{(2\pi)^{3}}\,W_{\ell}^{r}(k) \, , 
\ee
with
\be
	W_{\ell}^{r}(k) = \frac{\beta}{k}\int\,{\rm d}\chi\frac{{\rm d}N}{{\rm d}\chi}\,j'_{\ell}(k\chi) \, , 
\ee
where $\beta$ is the growth parameter defined by $\beta \equiv 1/b_q $ ${\rm (d}\ln D/{\rm d}\ln a) $, and 
$j'_{\ell}$ is the derivative of the spherical Bessel function with respect to its argument. We can then apply the fact that $C_{\ell} \equiv \langle g_{\ell} \, g_{\ell}^{*} \rangle$, and calculate the redshift-space-distorted angular power spectrum. 

\subsection{Non-linearities}
\label{sec:non_linear}

Non-linearities in the power spectrum are caused by the non-linear evolution of components of the Universe, especially the late-time evolution of matter and baryons. To capture the full extent of the non-linearities, with a lack of full-fledged non-linear evolution theory, one would need to simulate the evolution of most, if not all, of the components of the Universe. Extensive research into this problem has been conducted on multiple fronts \citep{crocce06, crocce08, sanchez08,sanchez09}, whether by perturbation theory \citep{crocce06, crocce08, carlson09}, dark matter simulations \citep{hamaus10, heitmann09}, or fitting functions suggested by dark matter simulations \citep{smith03}. Historically, there are a few ways to deal with non-linearities in using power spectra to constrain cosmology: 1) comparing the non-linear power spectrum to the linear power spectrum (usually for a specific cosmological model), and keeping only scales that are believed to be linear \citep{tegmark04, padmanabhan07}; 2) utilizing the halo occupation model to convert a quasar power spectrum into a halo power spectrum, which can easily be compared to halo power spectra from dark matter simulations \citep{reid10}; 3) using a variety of fitting functions developed \cite{carlson09} to match observed quasar power spectra \citep{blake10}. Our approach of relying on imaging data to derive angular power spectra is a benefit and a drawback. On the one hand, any line-of-sight signals are smeared out, as we lack  precise redshifts; but this also means we do not have to model the redshift space distortions in much details. 
Traditionally angular power spectrum analysis usually only applies a simple cut on the angular scale that roughly corresponds to $k= 0.1 h \ \rm{Mpc}^{-1}$ \citep{padmanabhan07}. In this paper, we follow \cite{ho12} and take a small step forward in terms of treating the overall shape of the angular power spectrum in the non-linear context. We adopt the simple linear redshift-independent biasing model --- with shot noise subtracted for each angular power spectrum. Therefore, in addition to the cosmological parameters that are of
interest for each model, we include three extra parameters for each redshift slice ($b$, $N_{\rm{shot}}$ and $a$) as shown in Eq.\ (\ref{eq:non-linear_eq}). A detailed test of this model can be found in \cite{ho12}. 

\subsection{Optimal Estimation of the Angular Power Spectrum}
\label{sec:est_theory}

The theory behind optimal power spectrum estimation is now well-established, so we limit ourselves to details specific to this discussion, and refer the reader to the numerous references on the subject \citep[][and references therein]{hamilton97,seljak98, ho08, ho12}. We also refer the reader to the Appendix in \cite{ho12} for more specific details that directly relate to our work.

We start by parameterizing the power spectrum with twenty step functions in $l$, $\tilde{C}^{i}_{l}$,
\be
	C_{\ell} = \sum_{i} p_{i} \tilde{C}^{i}_{\ell} \, ,
\ee
where the $p_{i}$ are the parameters that determine the power spectrum. We form quadratic combinations of the data,
\be
	q_{i} = \frac{1}{2} {\mathbf x}^{T} {\mathbf C}_{i} {\mathbf C}^{-1} {\mathbf C}_{i} {\mathbf x} \, ,
\ee
where ${\mathbf x}$ is a vector of pixelized quasar overdensities, ${\mathbf C}$ is the covariance matrix of the data, and ${\mathbf C}_{i}$ is the derivative of the covariance matrix with respect to $p_{i}$. The covariance matrix requires a prior power spectrum to account for cosmic variance; we estimate the prior by computing an estimate of the power spectrum with a flat prior and then iterating once. We also construct the Fisher matrix,
\be
	F_{ij} = \frac{1}{2}
{\rm tr} \left[{\mathbf C}_{i} {\mathbf C}^{-1} {\mathbf C}_{j}
{\mathbf C}^{-1}\right].
\ee
The power spectrum can then be estimated as $\hat{\mathbf p} = {\mathbf F}^{-1} {\mathbf q}$, with covariance matrix ${\mathbf F}^{-1}$.

\subsection{The Optimally Estimated Angular Power Spectrum}
\label{sec:angpower}

\begin{figure}
\begin{center}
	\includegraphics[width=2.75in,angle=0]{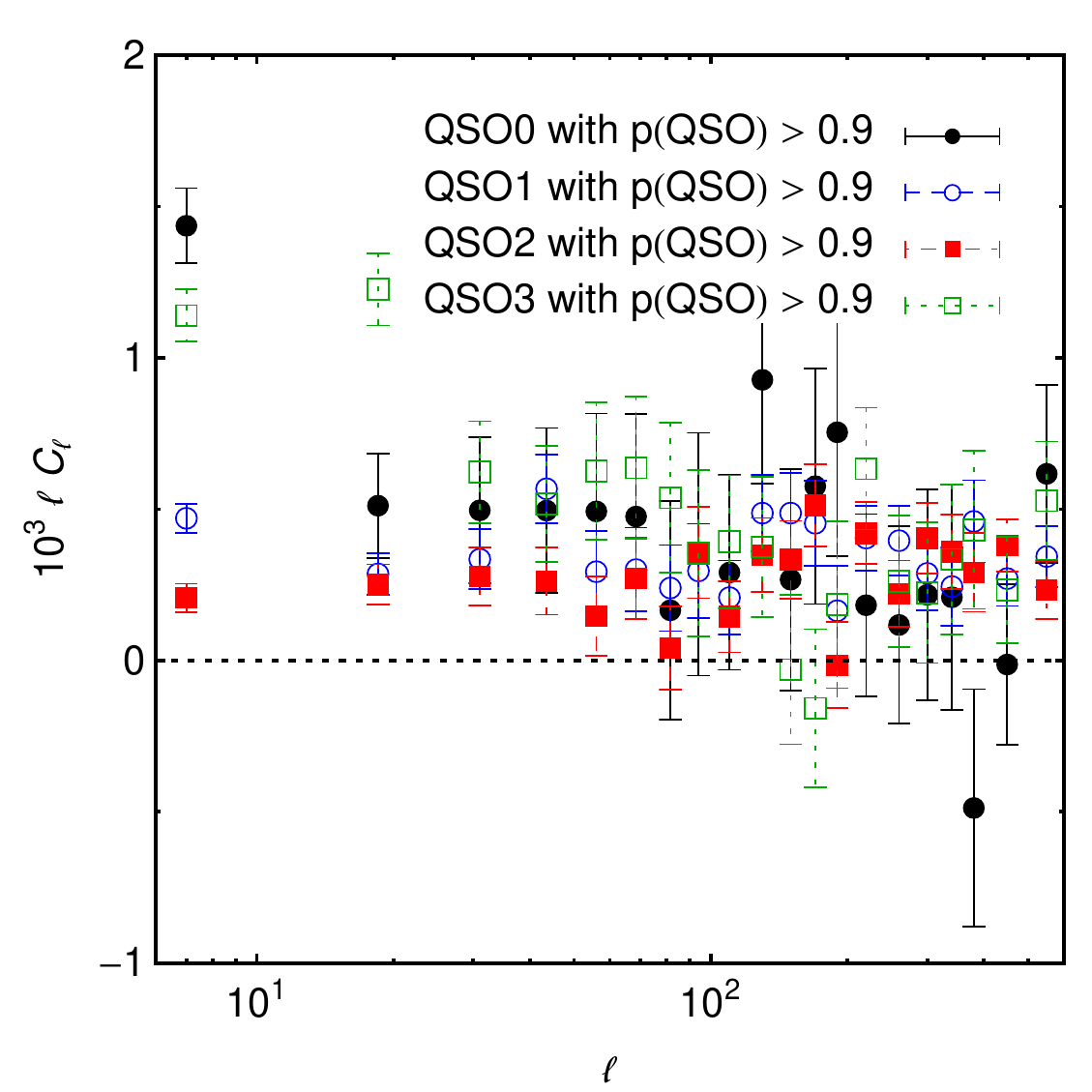}
	\caption{The measured angular power spectrum of quasars in our four redshift bins, obtained using the methodology described in \S\ref{sec:est_theory}. This does not include any of the corrections that we will apply later.}
\label{fig:cl2dfine}
\end{center}
\end{figure}

The angular power spectra of photometric quasars as described in \S\ref{sec:qsodef} and estimated using the methodology described in \S\ref{sec:est_theory} are displayed in Fig.\ \ref{fig:cl2dfine}. As shown in Fig.\ \ref{fig:dndz}, we must  investigate the potential effects of overlapping redshift distributions. Cross-power between different redshift bins not only adds cosmological information, but also information on systematics. When we examine cross-power across various redshift bins, any difference between the measured power and the expected power (from quasar auto-correlations in the same redshift range) can constrain systematics. In particular, we refer readers to our companion paper \cite{agarwal13} for a discussion of the use of cross-correlations to characterize unknown systematics.

\section{Potential Sample Systematics}
\label{sec:sys}

When using an imaging survey to infer power spectra, it is extremely important to account for any potential systematics that could affect the observed number density of objects in a given sample. Without properly correcting for sources of contamination we cannot claim an accurate measurement of the angular power spectrum and hence cannot use it to extract cosmological information. Below we review five known sources of contamination in the data that may contribute to extra (or deficit) power on the angular scales under consideration in this work, namely; stellar obscuration, sky brightness, seeing variations, dust extinction, and color offsets. We will show that only the first three of these potential contaminants are significant for quasar power spectra. The density maps and auto-power spectra for these systematics (using the imaging mask of Fig.\ \ref{fig:qsomask}) can be found in \cite{ho12}. In the next section we will discuss our method to correct quasar auto-power spectra for various systematics.

\subsection{Stellar Obscuration} 

\begin{figure}
\begin{center}
	\includegraphics[width=2.75in,angle=0]{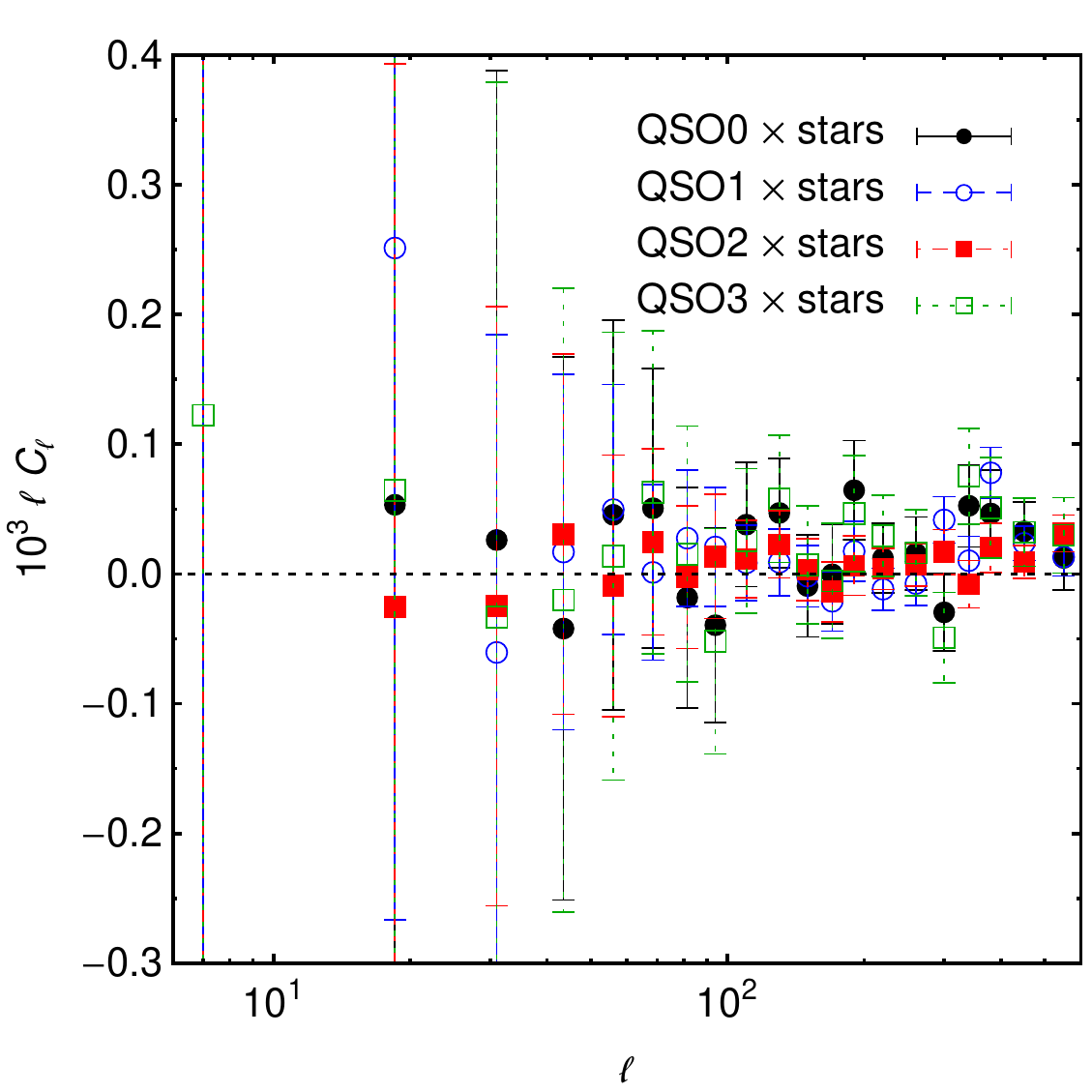}
	\caption{The cross-correlations between stellar overdensities in the range $18<r<18.5$ and  quasar overdsensities in our four redshift slices. There is surprisingly no significant contamination on large scales by stars for most redshift slices at scales $\ell \sim 10$, but there are significant contaminations on smaller scales such as $\ell \sim 100-200$.}
\label{fig:qsostarcross}
\end{center}
\end{figure}

Stars produce two types of contaminations in the creation of photometrically classified samples of quasars. As the color loci of quasars and stars in the SDSS intersect, stars can mimic quasars in certain redshift ranges. 
In addition, the light from bright foreground objects such as stars affects sky subtraction and can lead to contamination in the observed sample.  We evaluate these effects by investigating the cross-correlations between stars and quasars. We pick stars in the magnitude range $18 < r < 18.5$ and investigate their effects on quasar power spectra, applying the same imaging masks to both quasars and stars. Fig.\ \ref{fig:qsostarcross} displays the cross-correlations between quasars in different redshift slices and our sample of stars. 
From previous studies in \citep{ho08, hirata08}, we saw that stellar contamination of UVX objects is fairly common, which will give expected extra low-$\ell$ power when we cross-correlate stars with the photometric quasar catalog, as stars lie in our galactic plane, giving rise to extra power  at large scales. 
There is surprisingly no significant contamination on large scales by stars for most redshift slices at scales $\ell \sim 10$ , but there are significant contaminations on smaller scales such as $\ell \sim 100-200$. It is not obvious why this is the case, given our previous argument, but with our method of systematics removal, stellar obscuration will not affect the cosmological interpretation of our result. 

\subsection{Seeing Variations}

\begin{figure}
\begin{center}
	\includegraphics[width=2.75in,angle=0]{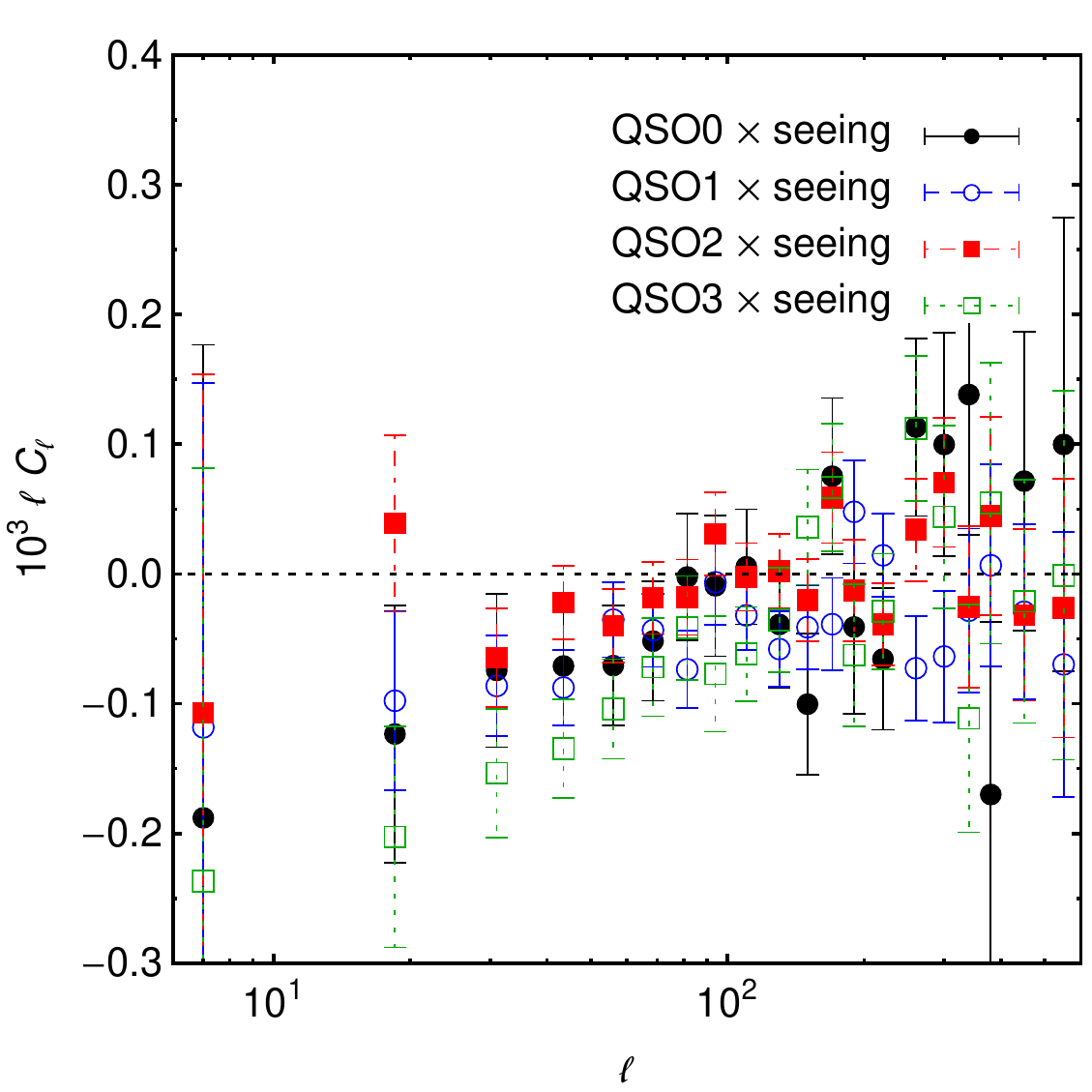}
	\caption{Cross correlations between image quality (seeing) and quasar overdensities in our four redshift slices.  We observe that there is a systematic and significant anti-correlation over nearly all scales relevant to the analysis between seeing and quasar overdensities. } 
\label{fig:qsoseeingcross}
\end{center}
\end{figure}

Since the SDSS uses a ground-based telescope, it is expected that the image quality, which is primarily degraded because of atmospheric seeing, will affect the number of quasars detected in any part of the sky. In fact, we find that seeing is the {\em most dominant} systematic for quasars. There is a systematic and significant anti-correlation over nearly all scales relevant to the analysis between seeing and quasar overdensities. On fitting the correlations (the angular auto-power spectrum of quasars vs. that of seeing, with $\ell \geq 30$) at various redshifts with a linear fit, we found the slope to be $-0.37 \pm 0.17$ at $z = 0.75$, $-0.20 \pm 0.14$ at $z = 1.25$, $-0.12 \pm 0.09$ at $z = 1.75$, and $-0.29 \pm 0.22$ at $z = 2.25$. This result is not entirely surprising as bad seeing\footnote{We already removed extremely bad seeing ($> 2.0''$ FWHM) areas early on in the analysis.} directly affects our ability to identify point sources. In \S\ref{sec:theory_sys} we use the cross-correlations between quasar overdensities and seeing variations shown in Fig.\ \ref{fig:qsoseeingcross} to determine, and remove, the effects of seeing on observations of quasar clustering.

\subsection{Sky Brightness}

\begin{figure}
\begin{center}
	\includegraphics[width=2.75in,angle=0]{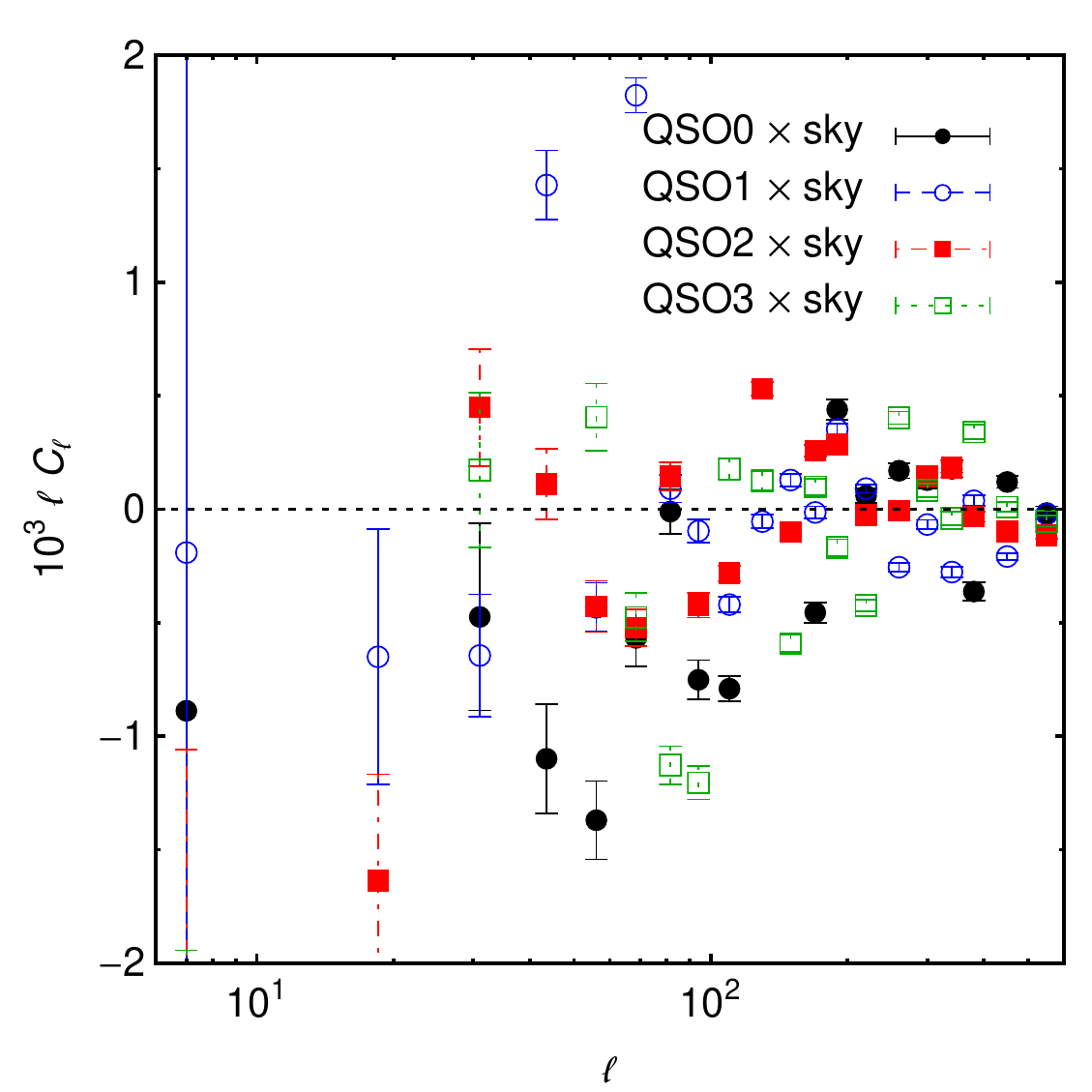}
	\caption{The cross-correlations between sky brightness (in the $i$-band) and quasar overdensities in our four redshift slices. Although there are significant cross-correlations between sky brightness and the quasar density, the effect on quasar overdensity scales as $\langle {\rm QSO0} \times {\rm SKY} \rangle /  \langle \rm{SKY}  \times \rm{ SKY} \rangle  $. Since the sky auto-correlations are relatively strong \protect \citep{ho12}, the actual effect on quasar overdensity is actually quite small.}
\label{fig:qsoskycross}
\end{center}
\end{figure}

The sky signal is subtracted from SDSS imaging scans before we use XDQSOz to identify quasars. 
There is a strong systematic effect with celestial location as when one moves to more southern declinations, one generally observes at higher zenith angles, so, in particular, imaging in the SDSS southern cap suffers from a brighter sky. Fig.\ \ref{fig:qsoskycross} displays the cross-correlations between photometrically classified quasars in different redshift ranges and the sky. Although there are significant cross-correlations between sky brightness and the quasar density, the effect on quasar overdensity scales as $\langle {\rm QSO0} \times  {\rm SKY} \rangle /  \langle \rm{SKY}  \times \rm{ SKY} \rangle  $. Since the sky auto-correlations are relatively strong \citep{ho12}, the actual effect on quasar overdensity is actually quite small. We also don't see a systematic trend in the cross-correlations. 

\subsection{Dust Extinction}

\begin{figure}
\begin{center}
	\includegraphics[width=2.75in,angle=0]{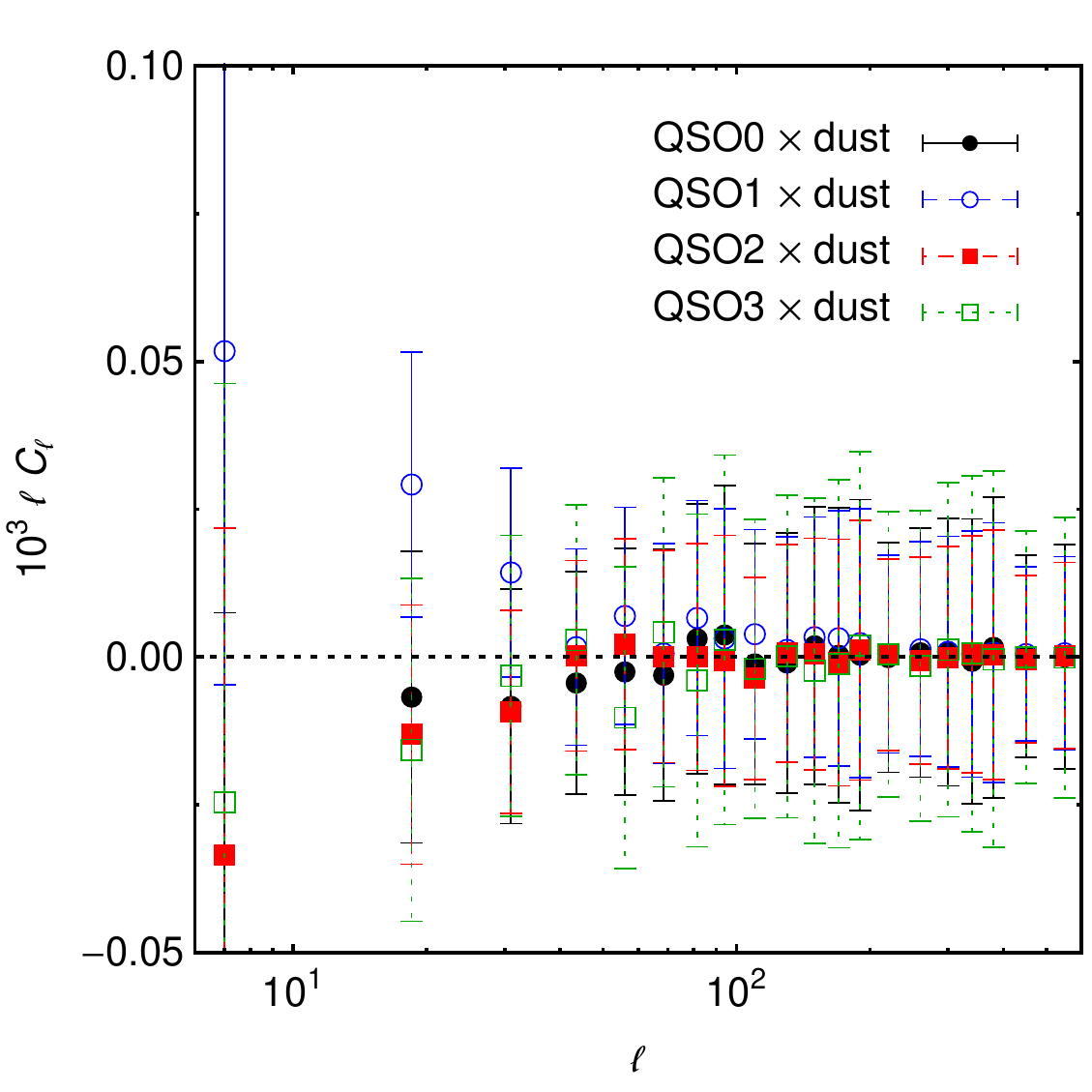}
	\caption{The cross-correlations between the Galactic extinction map and quasar overdensities in our four redshift slices. Since there are no significant correlations between quasar overdensities and extinction, we discard dust as a critical systematic. Some redshift slices have higher contamination at low multipoles, but since we exclude these extremely low multipoles in our analysis, dust extinction should not affect our inferred cosmological results.}
\label{fig:qsodustcross}
\end{center}
\end{figure}

We check for any residual effects of Galactic extinction on the observed overdensity of quasars by computing the cross-correlation of our quasars in different redshift ranges with the extinction map of \cite{schlegel98}; these cross-correlations are displayed in Fig.\ \ref{fig:qsodustcross}. Since the SDSS avoids areas with heavy dust extinction, we only have a small overlapping area where there is significant extinction and we  do not expect to see a statistically significant cross-correlation. Fig.\ \ref{fig:qsodustcross} does not reveal any statistically significant cross-correlation between our quasar and extinction maps, so we discard Galactic extinction as a possible contaminating systematic. 

\subsection{Color Offsets}

\begin{figure*}
\begin{center}
\leavevmode
	\includegraphics[width=2.75in,angle=0]{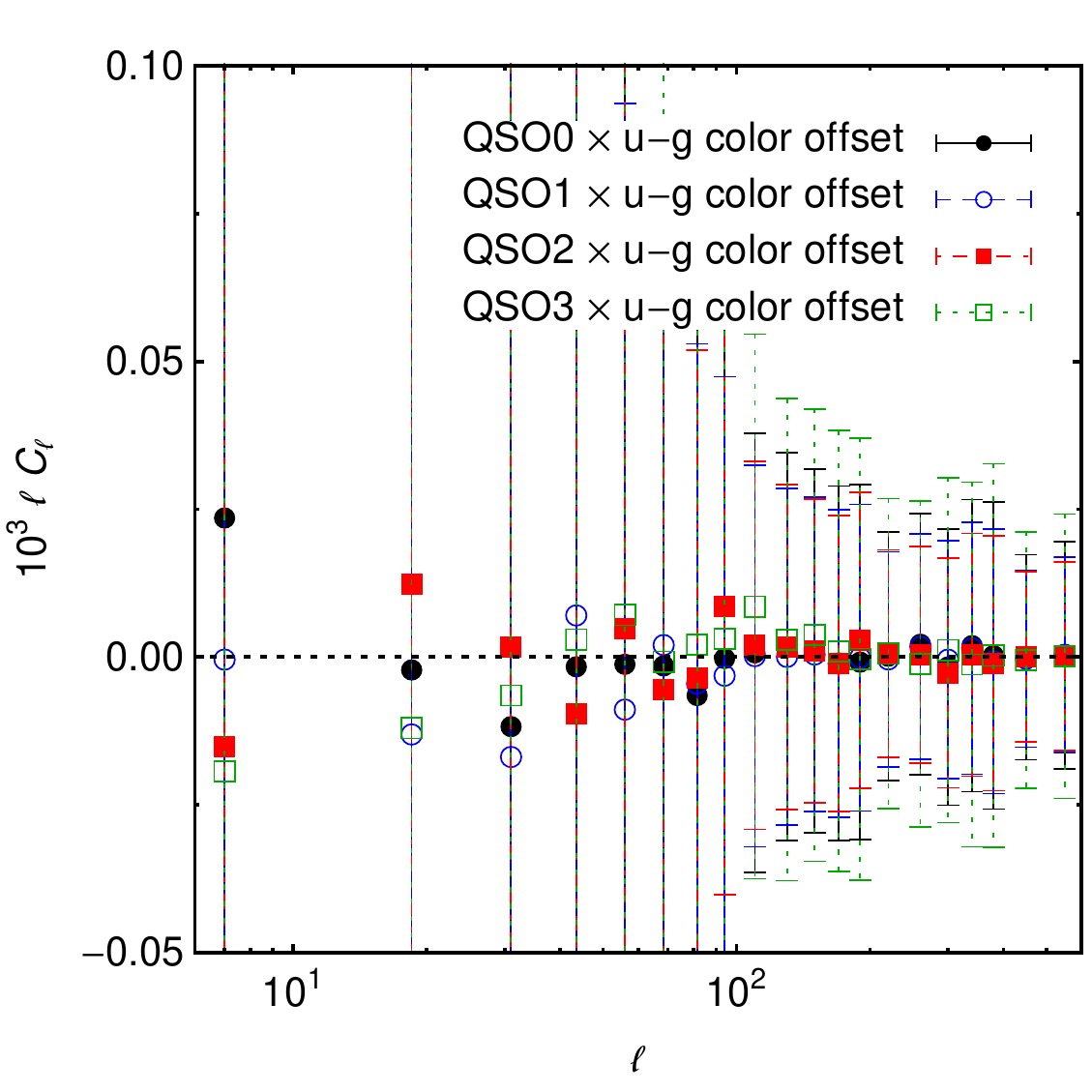}
	\includegraphics[width=2.75in,angle=0]{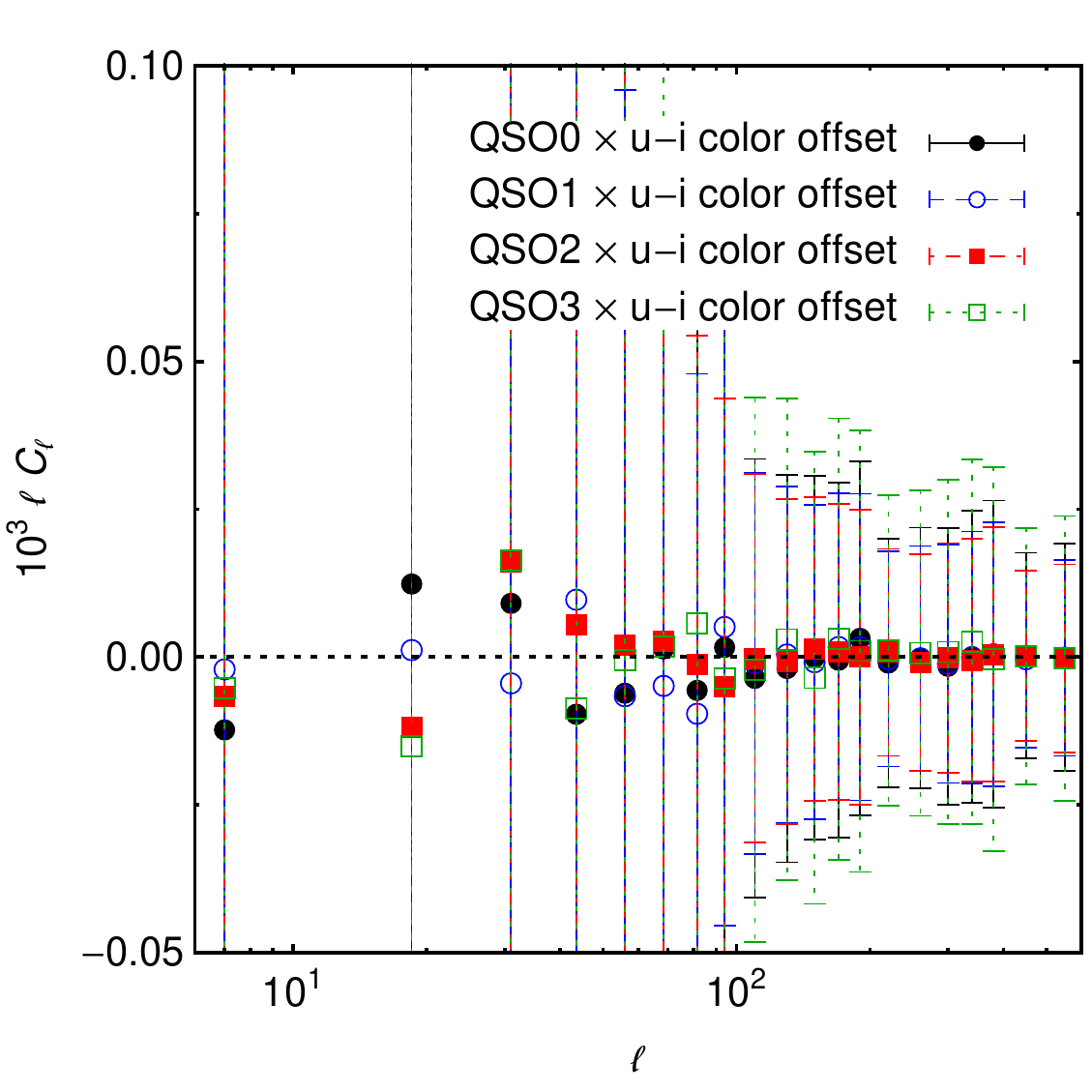}
	\includegraphics[width=2.75in,angle=0]{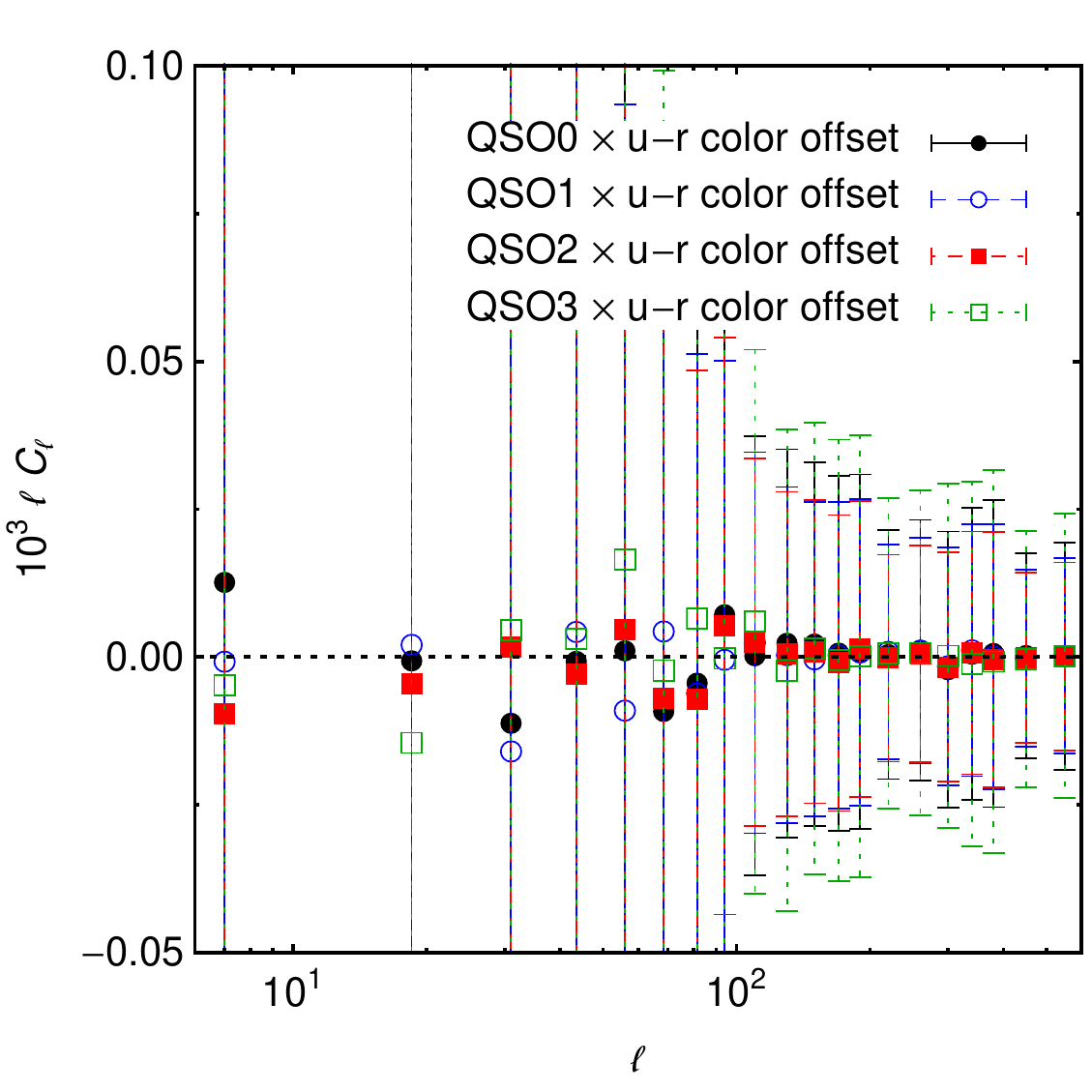}
	\includegraphics[width=2.75in,angle=0]{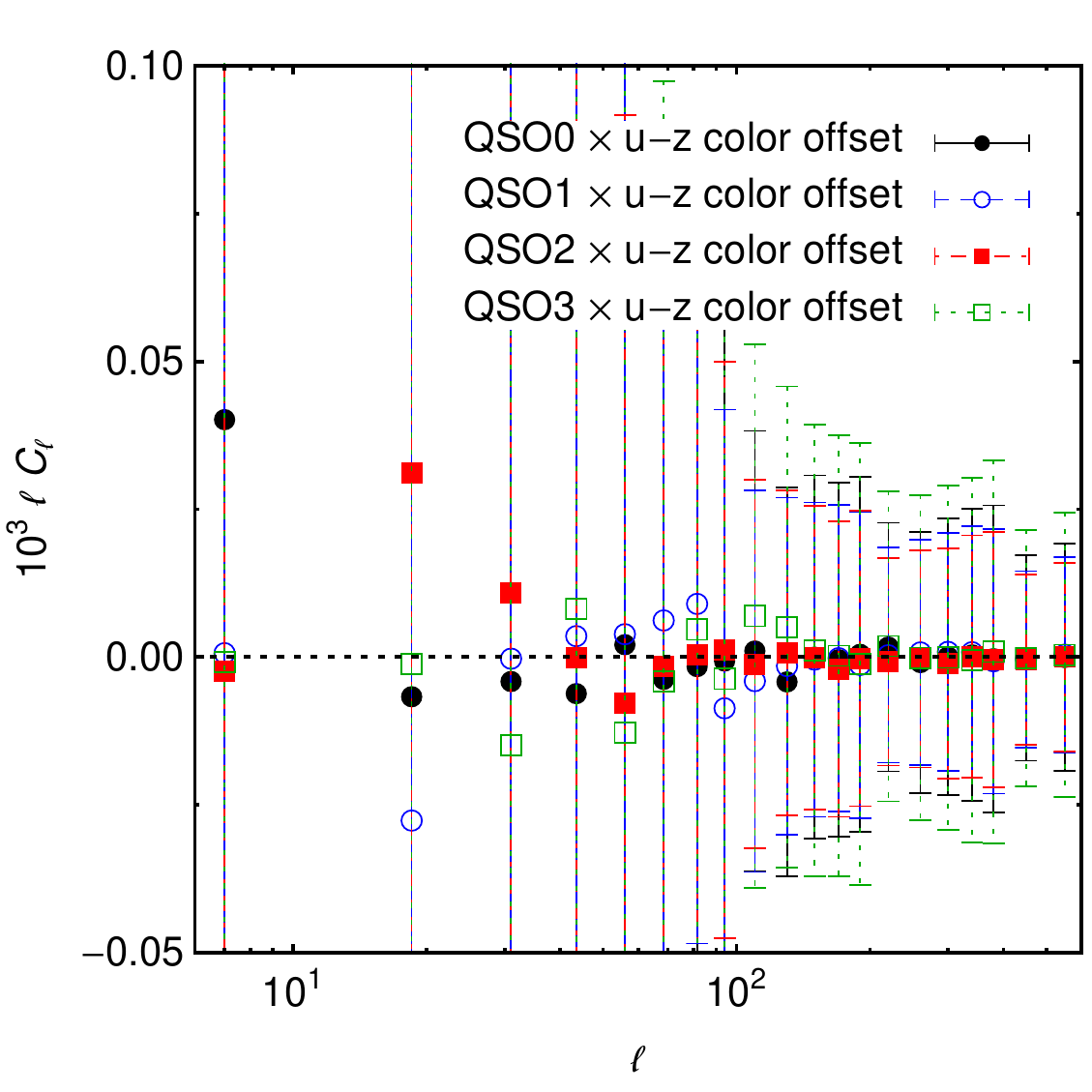}
	\caption{The cross-correlations between color offsets in $u-g$, $u-i$, $u-r$, and $u-z$, and quasar overdensities in our four redshift slices. There is no significant correlation between quasars and the color offsets, so we discard these offsets as a potential systematic.}
\label{fig:qsocolorcross}
\end{center}
\end{figure*}

To calibrate the measured magnitude of objects in the sky requires extreme care, and it is a topic that requires a long list of references. For example, in SDSS, we use a method called ``ubercal" \citep{padmanabhan07} which uses 
overlapping observations to solve for the calibration parameters and relative stellar fluxes simultaneously.  The relative photometric calibration in SDSS is accurate to $\sim 1\%$ after this procedure. 
This was further investigated recently by \citep{schlafly10}, and they found by observing the difference between observed and expected colors of stars in SDSS an offset in the photometric calibration, which exists in the current SDSS imaging dataset.  
The expected colors are found by inferring the spectral classes of stars from their spectra. 
We therefore test if there is any correlation between quasars and the offsets in all 4 colors. 
We found no significant correlation between overlapping regions of the quasar density maps and maps of the offsets, and hence conclude that color offsets are not a significant source of contamination in the quasar density fields (see  Fig.\ \ref{fig:qsocolorcross}). 

\section{Removing Systematics}
\label{sec:theory_sys}

The observed quasar density fields in different redshift slices must  be corrected for any sources of contamination before using their power spectra for cosmology. As discussed in the previous section, the most dominant systematic fields are those of seeing, stars and the sky. To correct for these systematics we adopt the method discussed in \cite{ross11, ho12}. Following \cite{agarwal13} we also use cross-correlations between quasars in different redshift slices to characterize the level of unknown contamination in the auto-power spectra of quasars. We exclude  from our analysis any $\ell$ bins that appear to be significantly contaminated with unknown systematics. Here we will briefly review this method and refer the interested reader to \cite{ross11, ho12, agarwal13} for details.

We adopt a simple linear relationship between $N_{\rm sys}$ systematics and the observed quasar density field in redshift slice $\alpha$,\footnote{For discussions on more general multiplicative errors see \cite{Gordon:2005ai,Huterer:2005ez,Huterer:2012zs,Hernandez-Monteagudo:2013vwa,Anderson:2013zyy}. For alternate methods of dealing with systematics see, e.g., \cite{Huterer:2012zs,Pullen:2012rd,Giannantonio:2013uqa,Hernandez-Monteagudo:2013vwa,Leistedt:2013gfa,Leistedt:2014wia,Shafer:2014wba}.}
\bea
	\delta_{q,{\rm obs}}^{\alpha}(\ell,m) & = & \delta_{q,{\rm true}}^{\alpha}(\ell,m) + \sum_{i=1}^{N_{\rm sys}} \epsilon_{i}^{\alpha}(\ell) \delta_{i}(\ell,m) + u^{\alpha}(\ell,m) \, . \quad
\eea
Here $\delta_{q,{\rm obs}}^{\alpha}(\ell,m)$ and $\delta_{q,{\rm true}}^{\alpha}(\ell,m)$ are the observed and true quasar density fields, $\delta_{i}(\ell,m)$ refers to the $i^{\rm th}$ systematic, $\epsilon_{i}^{\alpha}(\ell)$ is a weight factor that characterizes the effect of the $i^{\rm th}$ systematic, and $u^{\alpha}(\ell,m)$ encodes any unknown contamination. The observed and true angular auto-power spectrum in each redshift slice is defined as $C_{\ell, {\rm obs}}^{\alpha,\alpha} \equiv \left\langle \delta_{q,{\rm obs}}^{\alpha}(\ell,m) \delta_{q,{\rm obs}}^{\alpha}(\ell,m) \right\rangle$ and $C_{\ell, {\rm true}}^{\alpha,\alpha} \equiv \left\langle \delta_{q,{\rm true}}^{\alpha}(\ell,m) \delta_{q,{\rm true}}^{\alpha}(\ell,m) \right\rangle$.

Assuming that the true density field is not correlated with any of the systematics, we can write the following set of $N_{\rm sys}$ equations in each $\ell$ bin of the $\alpha^{\rm th}$ redshift slice
\bea
	\left\langle \delta_{q,{\rm obs}}^{\alpha}(\ell,m) \delta_{j}(\ell,m) \right\rangle & = & \sum_{i = 1}^{N_{\rm sys}} \epsilon_{i}^{\alpha}(\ell) \langle \delta_{i}(\ell,m) \delta_{j}(\ell,m) \rangle + \langle u^{\alpha}(\ell,m) \delta_{j}(\ell,m) \rangle \, ,
\label{eq:solveep}
\eea
with $j = 1, \ldots, N_{\rm sys}$. The connection between the true and observed angular auto- or cross-power spectra is 
\bea
	C_{\ell, {\rm true}}^{\alpha,\beta} & = & C_{\ell, {\rm obs}}^{\alpha,\beta} - \sum_{i,j = 1}^{N_{\rm sys}} \epsilon_{i}^{\alpha}(\ell) \epsilon_{j}^{\beta}(\ell) \langle \delta_{i}(\ell,m) \delta_{j}(\ell,m) \rangle - U^{\alpha,\beta}_{\ell},
\label{eq:crosspower}
\eea
where $U^{\alpha,\beta}_{\ell}$ is the contribution from unknown systematics.

\begin{figure}[!h]
\begin{center}
	\includegraphics[width=6.0in,angle=0]{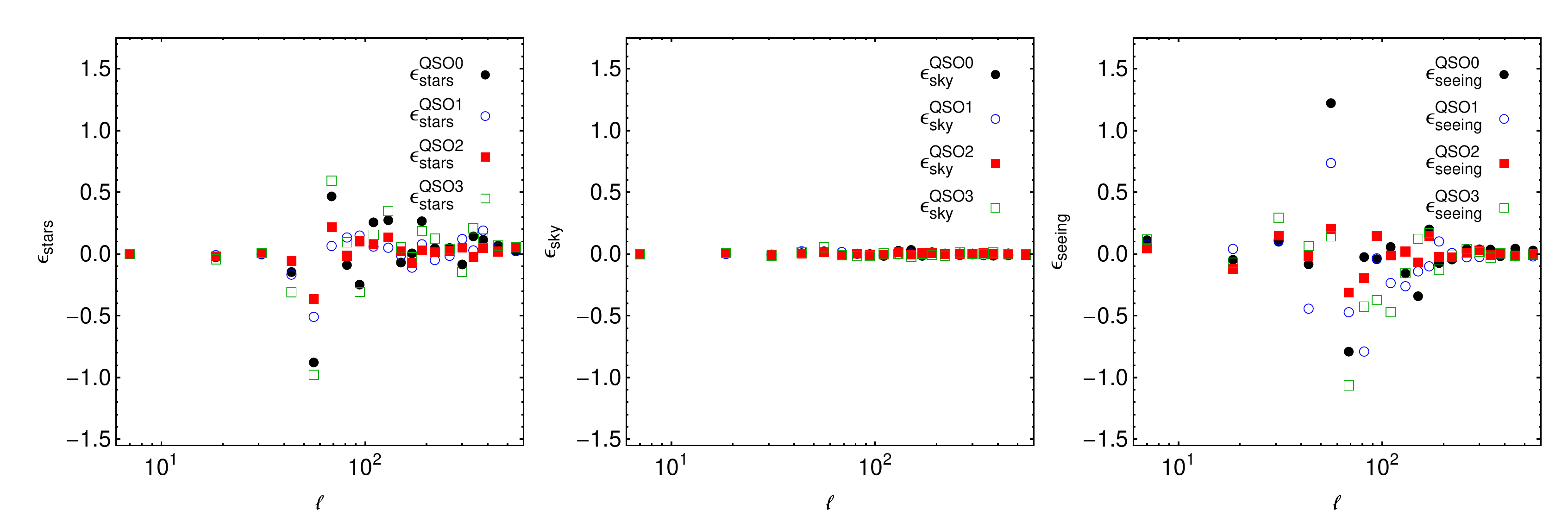}
	\caption{The weights $\epsilon_{i}^{\alpha}(\ell)$ obtained by solving Eq.\ (\ref{eq:solveep}) in each $\ell$ bin under the assumption that $u^{\alpha}(\ell,m) = 0$.}
\label{fig:epsilon}
\end{center}
\end{figure}

We first obtain a zeroth order estimate of the weights $\epsilon_{i}^{\alpha}(\ell)$ by solving Eq.\ (\ref{eq:solveep}) in each $\ell$ bin under the assumption that $u^{\alpha}(\ell,m) = 0$ (see Fig.\ \ref{fig:epsilon}). We then obtain the systematics-corrected auto-power spectrum in each redshift slice using Eq.\ (\ref{eq:crosspower}) for $\alpha = \beta$ and also assuming $U^{\alpha,\beta}_{\ell} = 0$. Before  these power spectra  can be used in an MCMC analysis, we also need to construct the full covariance matrix, which we will now describe.

\begin{figure}
\begin{center}
	\includegraphics[width=2.75in,angle=0]{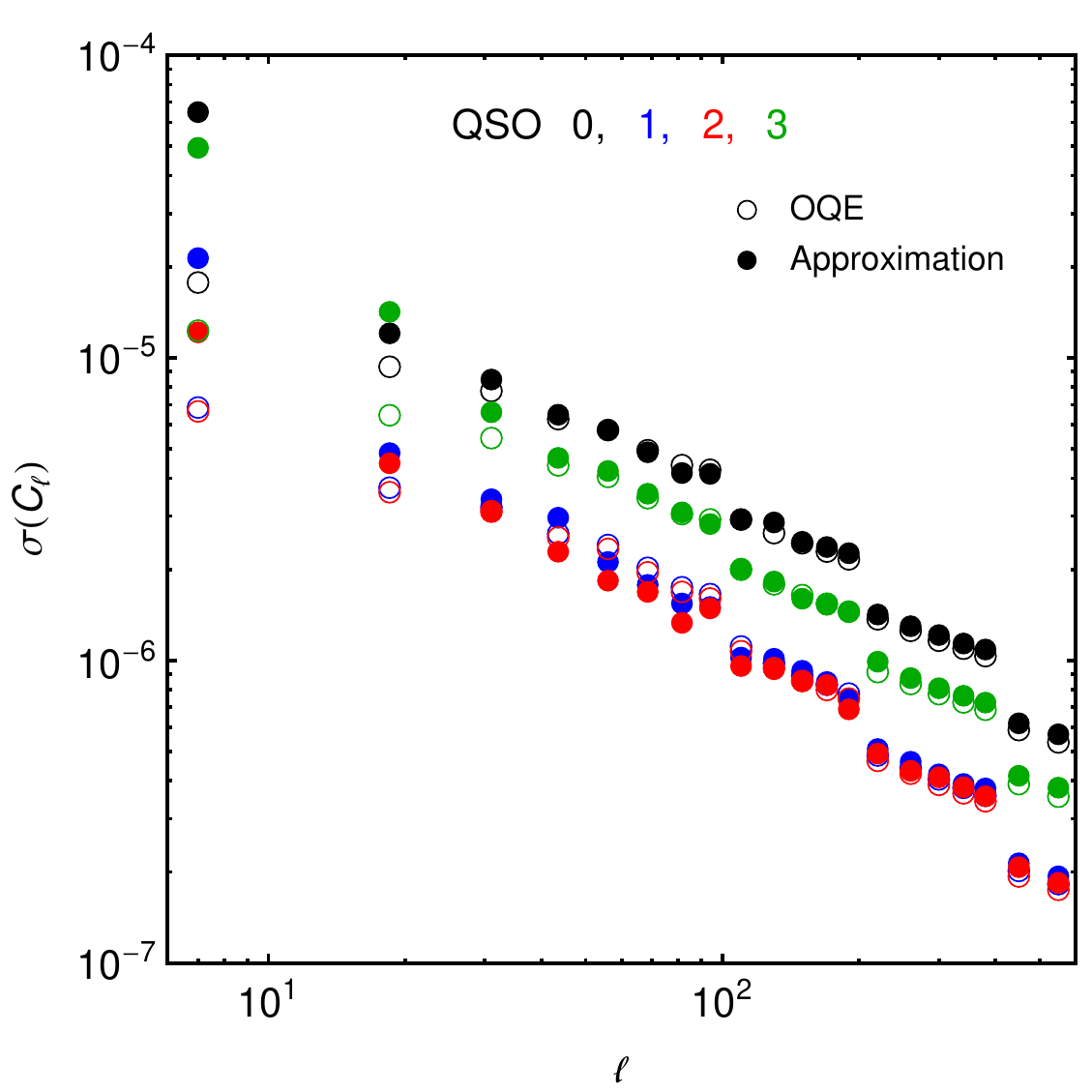}
	\includegraphics[width=2.75in,angle=0]{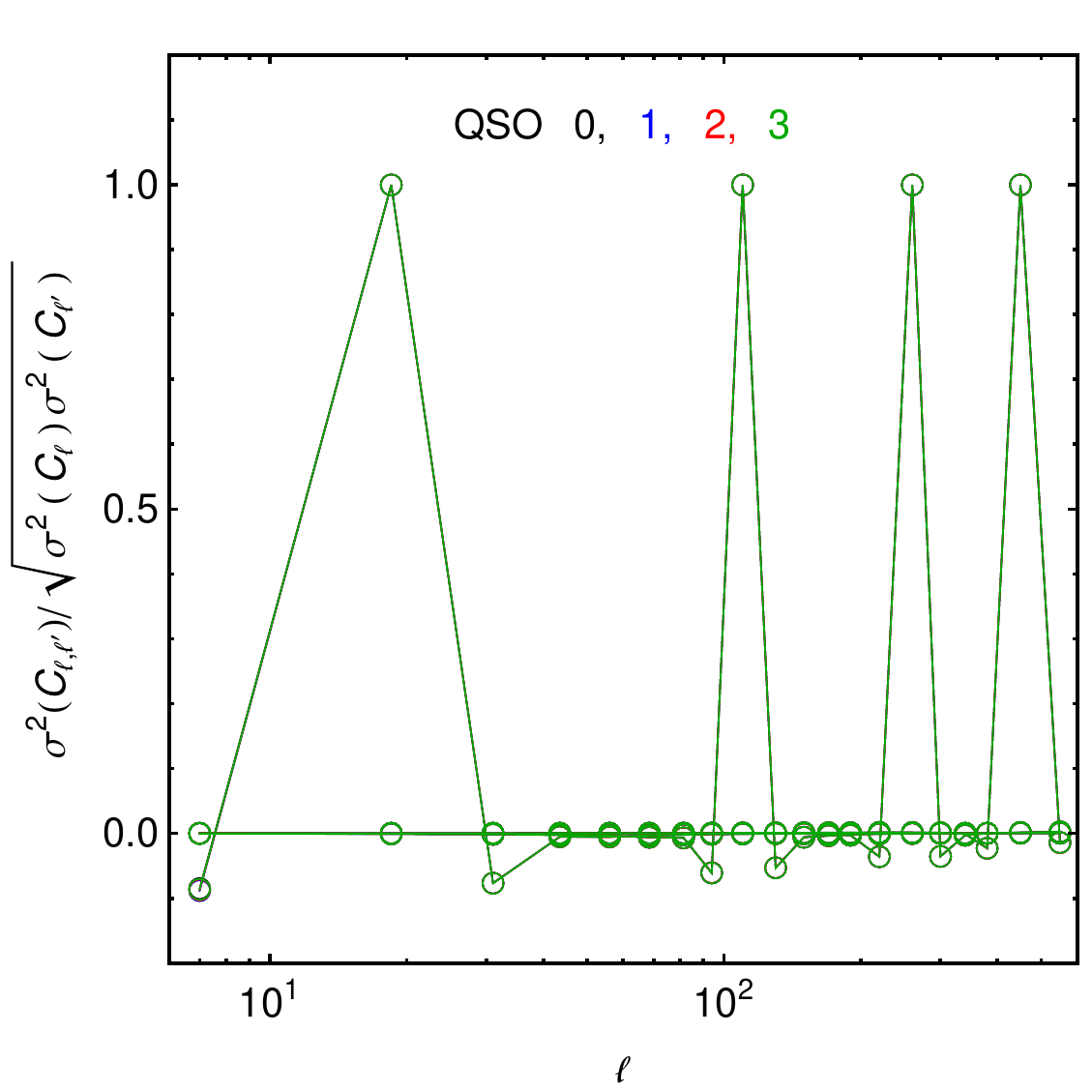}
	\caption{[LEFT] A comparison of the diagonal structure of the OQE covariance matrix and Eq.\ (\ref{eq:cov1}) in all four redshift slices. Open circles are the OQE prediction while closed circles are the modified Gaussian approximation applied to the systematics-corrected auto-power spectra. [RIGHT] Off-diagonal structure of the OQE covariance matrix in all four redshift slices. We show off-diagonal elements at four multipole slices: $\ell' = 18.5, \ 110.0, \ 260.0$, and $450.0$. The uniformity arises from a common mask.}
\label{fig:qsocovmatdiagonal}
\end{center}
\end{figure}

For the diagonal elements between redshift slices $\alpha$ and $\beta$ (i.e. for multipoles $\ell^{\alpha} = \ell^{\beta}$) we use a modified Gaussian approximation,
\bea
	\sigma^{2} \big( C_{\ell}^{\alpha,\alpha} \big) & = & a^{2}_{\rm fac} \frac{2}{f_{\rm sky} \sum_{\ell = \ell_{\rm min}}^{\ell_{\rm max} - 1}(2\ell + 1)} \left( \sqrt{ \left( C_{\ell,{\rm smooth}}^{\alpha,\alpha} \right)^{2} + \Big( \Delta C_{\ell}^{\alpha,\alpha} \Big)^{2} } + N_{{\rm shot},\alpha} \right)^{2}, \nonumber \\
\label{eq:cov1} \\
	\sigma^{2} \big( C_{\ell}^{\alpha,\alpha} C_{\ell}^{\beta,\beta} \big) & = & a^{2}_{\rm fac} \frac{2}{f_{\rm sky} \sum_{\ell = \ell_{\rm min}}^{\ell_{\rm max} - 1}(2\ell + 1)} \left( \left( C_{\ell,{\rm smooth}}^{\alpha,\beta} \right)^{2} + \left( \Delta C_{\ell}^{\alpha,\beta} \right)^{2} \right),
\label{eq:cov2}
\eea
where $f_{\rm sky}$ is the fraction of the sky observed, the shot noise $N_{\rm shot} = f_{\rm sky} \times 4\pi/N_{\rm sample}$, $N_{\rm sample}$ is the effective number of quasars observed, $\Delta C_{\ell}^{\alpha,\alpha} = \sum_{i,j = 1}^{N_{\rm sys}} \epsilon_{i}^{\alpha}(\ell) \epsilon_{j}^{\alpha}(\ell) \langle \delta_{i}(\ell,m) \delta_{j}(\ell,m) \rangle$ and $\Delta C_{\ell}^{\alpha,\beta} = \sum_{i,j = 1}^{N_{\rm sys}} \epsilon_{i}^{\alpha}(\ell) \epsilon_{j}^{\beta}(\ell) \langle \delta_{i}(\ell,m)$ $\delta_{j}(\ell,m) \rangle$.  This approach also takes into account the fact that the total number of modes in each $\ell$ bin is given by the sum of modes in $\ell_{\rm min} \leq \ell < \ell_{\rm max}$. Since the Gaussian approximation is not perfect and neighboring $\ell$ bins may contribute to the diagonal elements, the diagonal error can be boosted by an empirical factor of $a_{\rm fac}$. For quasars we do not need to boost the diagonal error and hence set this factor to unity. Fig.\ \ref{fig:qsocovmatdiagonal} compares the diagonal errors from the optimal quadratic estimator (henceforth OQE) and Eq.\ (\ref{eq:cov1}) in all four redshift slices. The two agree very well.

For the off-diagonal elements of the covariance matrix we simply preserve the structure of the OQE covariance matrix,
\bea
	\sigma^{2} \big( C_{\ell,\ell'}^{\alpha,\alpha} \big) & = & \frac{\sigma^{2} \big( C_{\ell,\ell',{\rm OQE}} \big)}{\sqrt{ \sigma^{2} \big( C_{\ell,{\rm OQE}} \big) \sigma^{2} \big( C_{\ell',{\rm OQE}} \big) }} \ \sqrt{\sigma^{2} \big( C_{\ell}^{\alpha,\alpha} \big) \sigma^{2} \big( C_{\ell'}^{\beta,\beta} \big)} \, ,
\label{eq:cov3} \\
	\sigma^{2} \big( C_{\ell}^{\alpha,\alpha} C_{\ell'}^{\beta,\beta} \big) & = & \frac{\sigma^{2} \big( C_{\ell,\ell',{\rm OQE}} \big)}{\sqrt{ \sigma^{2} \big( C_{\ell,{\rm OQE}} \big) \sigma^{2} \big( C_{\ell',{\rm OQE}} \big) }} \ \sqrt{\sigma^{2} \big( C_{\ell}^{\alpha,\alpha} C_{\ell}^{\beta,\beta} \big) \sigma^{2} \big( C_{\ell'}^{\alpha,\alpha} C_{\ell'}^{\beta,\beta} \big)} \, .
\label{eq:cov4}
\eea
All redshift slices usually have similar OQE covariance structures, as we show in Fig.\ \ref{fig:qsocovmatdiagonal}. We can therefore use any redshift slice to generate the ratio in the above equation.

\begin{figure*}
\begin{center}
\leavevmode
	\includegraphics[width=5.50in,angle=0]{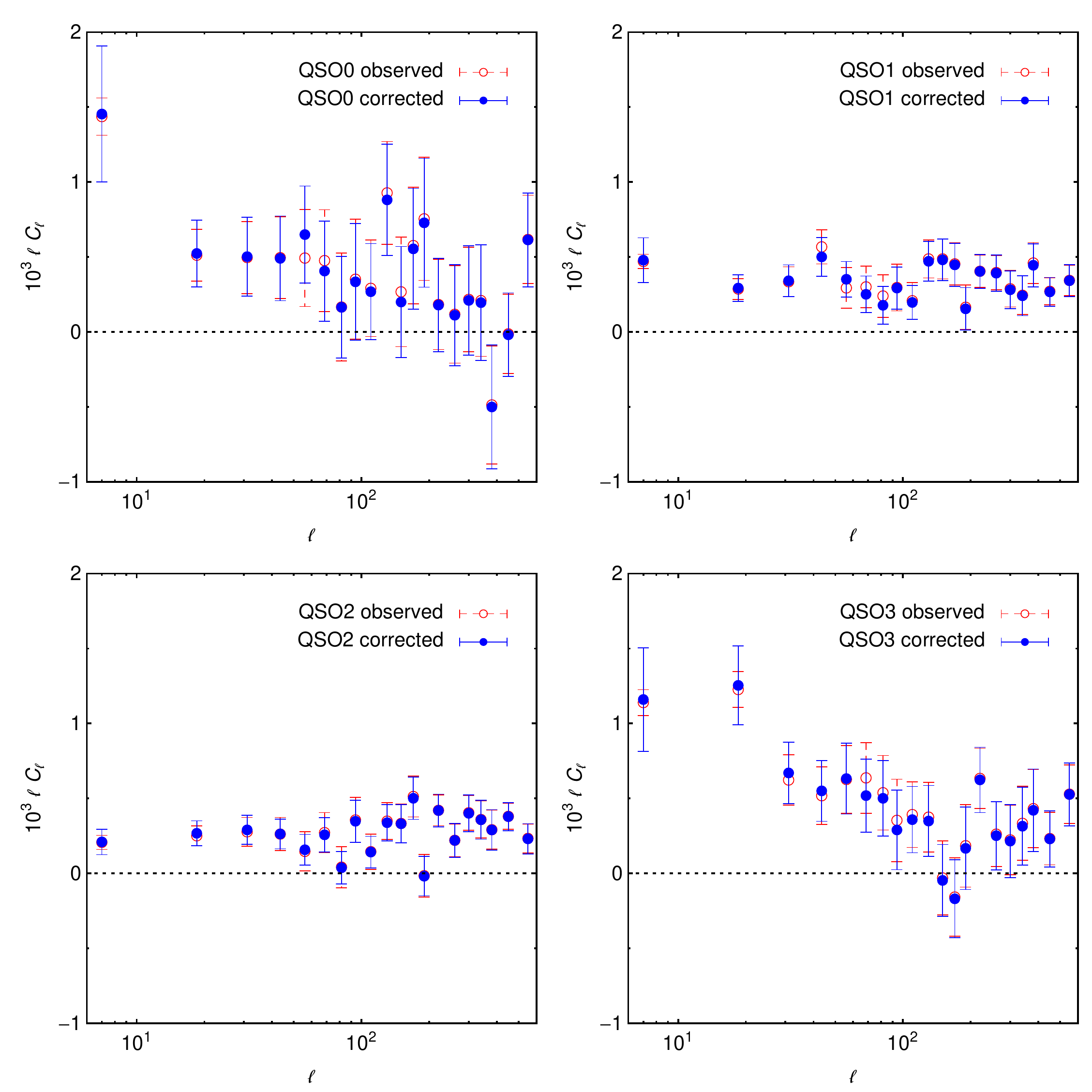}
	\caption{The measured and corrected angular power spectrum of quasars in our four redshift bins.}
\label{fig:qsocorrectedauto}
\end{center}
\end{figure*}

We show the measured (same as Fig.\ \ref{fig:cl2dfine}) and corrected auto-power spectra of quasars, with appropriate errors, in Fig.\ \ref{fig:qsocorrectedauto}. Using these power spectra in an MCMC analysis, we obtain the bias in each redshift slice.

\begin{figure*}
\begin{center}
	\includegraphics[width=5.50in,angle=0]{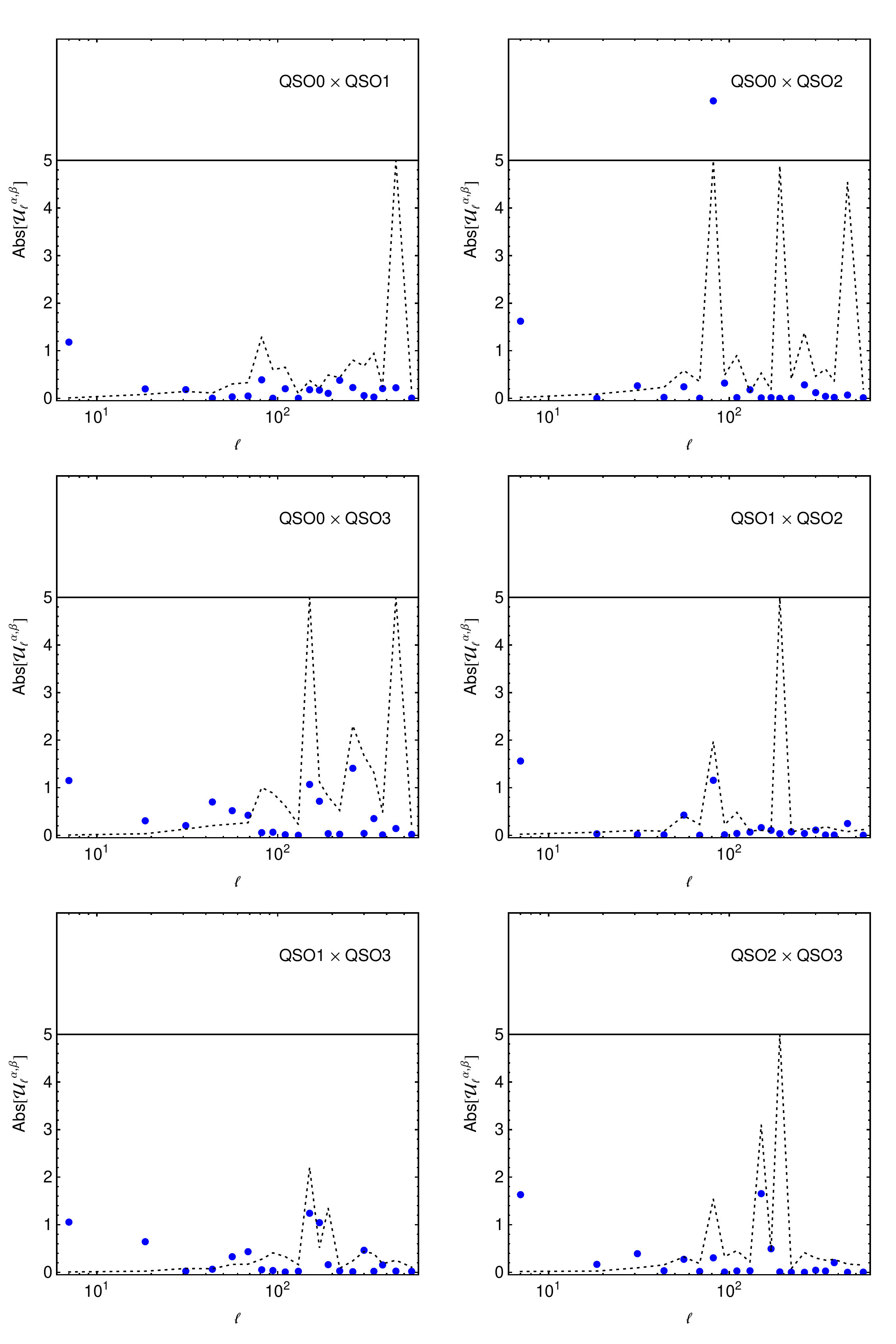}
	\caption{The absolute value of the unknown contamination coefficient ${\cal U}^{\alpha,\beta}_{\ell}$ for quasars (filled circles). The dotted line shows the absolute value of the $1\sigma$-cut --- we drop all bins that lie above this cut. The points that lie above the solid line have ${\rm Abs}\big[{\cal U}^{\alpha,\beta}_{\ell}\big] > 5.0$.}
\label{fig:qsoulul}
\end{center}
\end{figure*}

\begin{figure*}
\begin{center}
	\includegraphics[width=5.5in,angle=0]{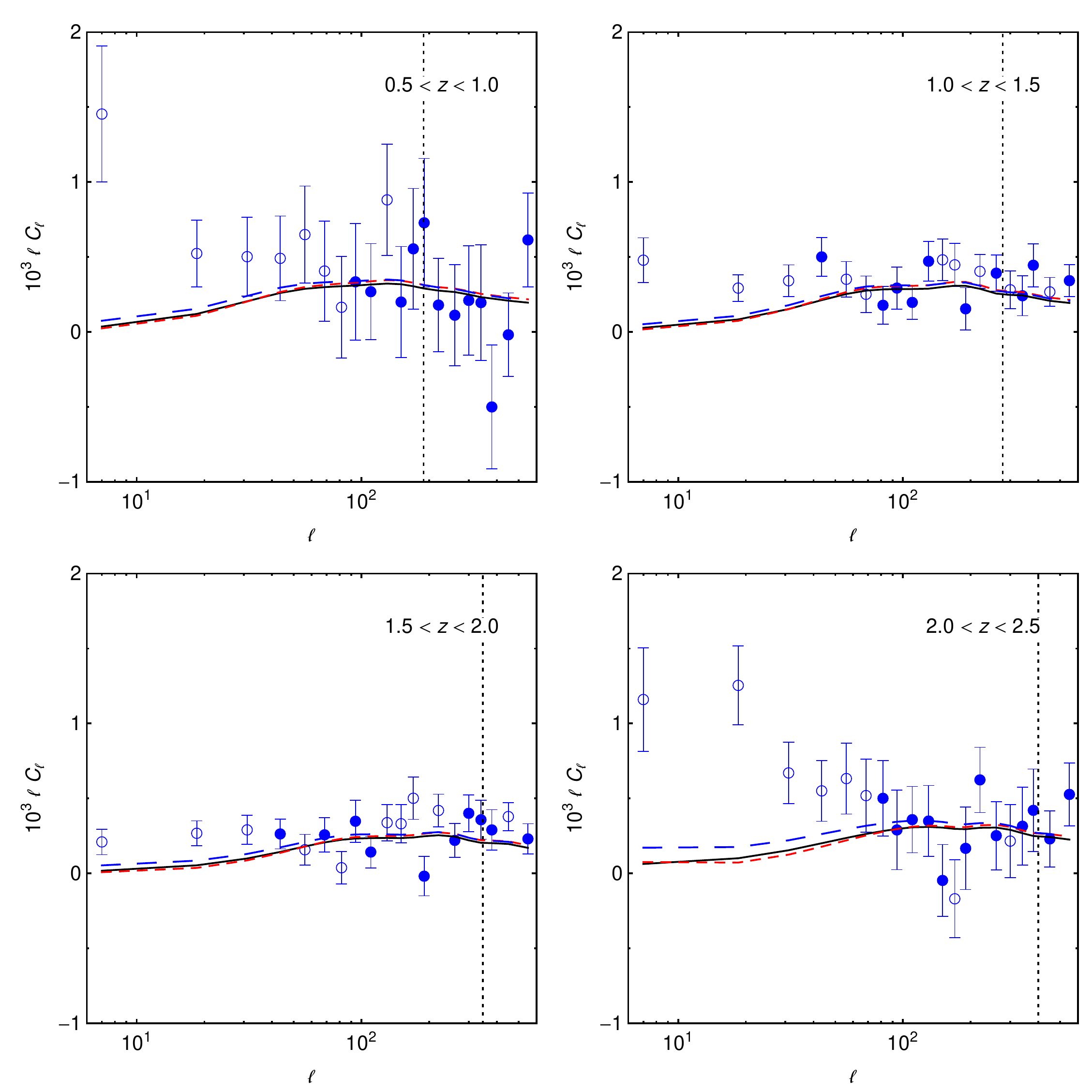}
	\caption{The corrected angular power spectrum of quasars in the four redshift slices. Open circles represent data points that are dropped due to large unknown systematics (final iteration). Filled circles are data points that are not dominated with unknown systematics. The vertical dotted line shows $\ell_{\rm max}$. The curves are the theoretical angular power spectra at the best-fit and $68\%$ confidence values of $f_{\rm NL}$ for quasars + LRGs --- $f_{\rm NL} = 2$ (solid black), $f_{\rm NL} = -64$ (dashed red), and $f_{\rm NL} = 67$ (long-dashed blue). For the theoretical angular power spectra, the Gaussian bias in each redshift slice is set to the value that corresponds to the MCMC analysis which only uses filled circles in $10 \leq \ell \leq \ell_{\rm max}$.}
\label{fig:qsocltheory}
\end{center}
\end{figure*}

For this purpose, we use a modified version of the widely used package {\tt CosmoMC} \cite{lewis02}. The  $\chi^{2}$ that is the input to the MCMC procedure is $({\bf d} - {\bf t})^{T} \ . \ {\bf {\cal C}}^{-1} \ . \ ({\bf  d} - {\bf  t})$, where ${\bf d}$ is the data $C_{\ell}$ vector, ${\bf t}$ is the theory $C_{\ell}$ vector of \S\ref{sec:theory1} convolved with the full window function, and ${\bf {\cal C}}$ is the covariance matrix. We calculate the linear matter power spectrum using the {\tt CAMB} code \cite{lewis00}, and apply the HaloFit prescription \cite{smith02} to account for non-linear effects on the matter power spectrum. We apply a low-$\ell$ cutoff on the angular power spectrum in each redshift slice at $\ell_{\rm min} = 30$ since we expect lower multipoles to mostly be dominated by systematics. We also choose a high-$\ell$ cutoff, $\ell_{\rm max}$, corresponding to $k = 0.1 h \ {\rm Mpc}^{-1}$ to avoid the strongly non-linear regime of the matter power spectrum. We also drop the extra non-linear fitting parameter $a$ (see Eq.\ (\ref{eq:non-linear_eq})) from our list of parameters to vary as the error bars on the data are too large to allow for a good fit to this parameter.

For the MCMC analysis we use standard cosmological data, including the WMAP nine-year CMB data \cite{hinshaw12,bennett12} and  the ``Union 2'' supernova data set that includes 557 supernovae \cite{amanullah10}, as our baseline model. In addition to the bias in each redshift slice, we also vary over the standard cosmological parameters $\big\{ \Omega_{b}h^{2}, \Omega_{\rm DM}h^{2}, \theta,\tau, n_{s}, \log A_{s}, A_{\rm SZ} \big\}$. Here $\Omega_{b}h^{2}$ is the physical baryon density, $\Omega_{\rm DM}h^{2}$ is the physical dark matter density, $\theta$ is the ratio of the sound horizon to the angular diameter distance at decoupling, $\tau$ is the reionization optical depth, $n_{s}$ is the scalar spectral index, $A_{s}$ is the amplitude of the primordial scalar curvature perturbations at $k =0.05 \ {\rm Mpc}^{-1}$, and $A_{\rm SZ}$ represents a Sunyaev-Zeldovich template normalization.

The MCMC analysis produces the bias values listed in the fourth column of Table\ \ref{table:qsobias} (this MCMC fit gives a $\Delta\chi^2$ per d.o.f. of $1.20$, relative to the $\Lambda$CDM fit using WMAP9 + SN data sets). These values of the bias are used to estimate the true cross-power spectrum between any two redshift slices. We obtain the cross-redshift distribution needed here as an overlap of the distribution in the two slices. Once we have the true cross-power, we compare it with the measured cross-power using Eq.\ (\ref{eq:crosspower}). Using the weights $\epsilon_{i}^{\alpha}(\ell)$ obtained earlier, this equation provides us with an estimate of $U_{\ell}^{\alpha,\beta}$ in each $\ell$ bin.\footnote{When calculating $U_{\ell}^{\alpha,\beta}$, we do not convolve the theoretical cross-power with the survey window function.}

We now calculate the unknown contamination coefficient defined in \cite{agarwal13} as 
\bea
	{\cal U}^{\alpha,\beta}_{\ell} & = & \frac{\left( U^{\alpha,\beta}_{\ell} \right)^{2}}{C_{\ell, {\rm obs}}^{\alpha,\alpha} C_{\ell, {\rm obs}}^{\beta,\beta}} \, , \quad \alpha \neq \beta \, ,
\label{eq:defineul}
\eea
and compare it to the quantity
\bea
	\frac{\sigma \big( C_{\ell, {\rm obs}}^{\alpha,\alpha} \big) \sigma \big( C_{\ell, {\rm obs}}^{\beta,\beta} \big)}{C_{\ell, {\rm obs}}^{\alpha,\alpha} C_{\ell, {\rm obs}}^{\beta,\beta}} \, .
\label{eq:usyssigma}
\eea
We discard all $\ell$ bins in each redshift slice $\alpha$ and $\beta$ for which ${\cal U}^{\alpha,\beta}_{\ell}$ is greater than the $1\sigma$-cut in Eq.\ (\ref{eq:usyssigma}). Using the remaining bins in $10 \leq \ell \leq \ell_{\rm max}$ and the corresponding rows and columns of the full covariance matrix obtained earlier, we perform an MCMC analysis on the standard cosmological parameters and the bias in each redshift slice. This generates the new bias values denoted as `$(1^{\rm st} \ {\rm it.})$' (First iteration) in Table\ \ref{table:qsobias} (with a $\Delta\chi^2$ per d.o.f. of $1.03$). This process is repeated a few times until the bias of each redshift slice lies within one sigma of the previous iteration. Table\ \ref{table:qsobias} lists the bias values obtained in the final iteration (with a $\Delta\chi^2$ per d.o.f. of $0.96$). Fig.\ \ref{fig:qsoulul} shows the absolute values of the unknown contamination coefficient and the $1\sigma$-cut for the final iteration. Finally, Fig.\ \ref{fig:qsocltheory} presents the corrected quasar auto-power spectrum in each redshift slice and marks the bins that are dropped on the basis of the method discussed here. We also show theoretical curves\footnote{The theoretical curves shown in Fig.\ \ref{fig:qsocltheory} differ from those used in calculating the likelihood by the effect of the survey window function.} obtained using best-fit values of the bias (with an exception for the first redshift slice) for $1\sigma$ constraints on $f_{\rm NL}$, which are discussed in the next section.

For the first redshift slice, after the final iteration, we use the {\it median} value of the bias, which is $b_{1} = 2.57$, instead of the {\it mean} value of $b_{1} = 2.19$. The reason for this choice is that the first redshift slice appears to be dominated by unknown systematics. As a result, using bins that survive the $1\sigma$-cut yields an estimate of the bias (which is varied in $b_{1} \in [0.1,10]$) that is not bounded from below (see Table\ \ref{table:qsobias}). We use the median value in all of the cosmological parameter analysis in the next section as well. The method used here excludes bins from {\it both} redshift slices whose cross-power is significantly contaminated as one cannot tell a priori which redshift slice is responsible for the contamination. For this reason, there will be bins in each redshift slice that do not appear contaminated but are still dropped as their cross-power with another redshift slice is significantly contaminated.

\begin{table}[!h]
\begin{center}
\begin{tabular}{|c|c|c|c|c|c|}
\hline
Label & $z_{\rm mid}$ & $l_{\rm max}$ & $b_{1}$ & $b_{1}$ & $b_{1}$ \\
& & & & $(1^{\rm st} \ {\rm it.})$ & $(2^{\rm nd} \ {\rm it.})$ \\
\hline
QSO0 & 0.75 & 189 & $3.01^{+0.18}_{-0.15}$ & $1.93^{+0.51}_{-1.83}$ & $2.19^{+0.47}_{-2.09}$ \\
QSO1 & 1.25 & 278 & $2.22^{+0.11}_{-0.11}$ & $2.14^{+0.07}_{-0.08}$ & $2.06^{+0.08}_{-0.08}$ \\
QSO2 & 1.75 & 346 & $2.45^{+0.14}_{-0.14}$ & $2.31^{+0.10}_{-0.09}$ & $2.32^{+0.11}_{-0.09}$ \\
QSO3 & 2.25 & 400 & $3.64^{+0.31}_{-0.31}$ & $3.43^{+0.21}_{-0.18}$ & $3.37^{+0.20}_{-0.18}$ \\
\hline
\end{tabular}
\caption{The best-fit Gaussian bias (with $1\sigma$ errors) for quasars in our four redshift slices, using WMAP9 + SN + DR8 (QSO) data. In the fourth column we use all available $\ell$ bins in $30 \leq \ell \leq \ell_{\rm max}$, while in the remaining columns we use only those bins in $10 \leq \ell \leq \ell_{\rm max}$ that satisfy a $1\sigma$-cut on ${\cal U}^{\alpha,\beta}_{\ell}$.}
\label{table:qsobias}
\end{center}
\end{table}

\section{Results}
\label{sec:results}

\subsection{Constraints on Cosmological Parameters}

The angular clustering of large scale structure allows us to constrain the background cosmological parameters that govern the evolution of our Universe. We use measurements of the quasar angular power spectrum in our four redshift slices QSO0--QSO3 to constrain cosmology in a flat $\Lambda$CDM model and to constrain primordial local non-Gaussianity.

We first present constraints for a flat $\Lambda$CDM model. On combining quasar angular power spectra with WMAP9 + SN data and performing an MCMC analysis over the standard cosmological parameters $\big\{ \Omega_{b}h^{2}, \Omega_{\rm DM}h^{2}, \theta, \tau, n_{s}, \log A_{s}, A_{\rm SZ} \big\}$ and the bias in all four redshift slices, we obtain the results shown in Table\ \ref{table:qsoresults}. Here $\Omega_{m}$ is the matter density in units of the critical density today, $\Omega_{\Lambda}$ is the dark energy density in the same units, and $\sigma_{8}$ is the RMS density fluctuation at $8 h^{-1}$ Mpc. Quasars do not significantly add constraining power to the vanilla cosmological model when compared to just WMAP9 + SN (first column). It is worth noting, however, that our method of removing contaminated bins using cross-correlations among different redshift slices (last column) yields slightly different results (in a direction consistent with constraints from WMAP9 + SN data alone) compared to the common choice of retaining $\ell$ bins in $30 \leq \ell \leq \ell_{\rm max}$ in quasar angular power spectra corrected for known systematics (middle column). The best-fit values of the bias were shown earlier in Table \ref{table:qsobias}.

\begin{table*}[!htb]
\begin{center}
	\begin{tabular}{|c|c|c|c|}
		\hline
		Parameter & WMAP9 + SN & WMAP9 + SN & WMAP9 + SN \\
		& & + QSO & + QSO \\
		& & (Before $1\sigma$-cut) & (After $1\sigma$-cut) \\
		\hline
		$\Omega_{b}h^{2}$ & $0.0228 \pm 0.0005$ & $0.0228 \pm 0.0005$ & $0.0228 \pm 0.0005$ \\
		$\Omega_{\rm DM}h^{2}$ & $0.112 \pm 0.004$ & $0.111 \pm 0.004$ & $0.112 \pm 0.004$ \\
		$\Omega_{m}$ & $0.274 \pm 0.021$ & $0.265 \pm 0.020$ & $0.272 \pm 0.020$ \\
		$\Omega_{\Lambda}$ & $0.726 \pm 0.021$ & $0.735 \pm 0.020$ & $0.728 \pm 0.020$ \\
		$h$ & $0.704 \pm 0.019$ & $0.711 \pm 0.019$ & $0.706 \pm 0.018$ \\
		$\sigma_{8}$ & $0.820 \pm 0.021$ & $0.812 \pm 0.022$ & $0.815 \pm 0.021$ \\
		$n_{s}$ & $0.973 \pm 0.012$ & $0.975 \pm 0.012$ & $0.973 \pm 0.012$ \\
		\hline
	\end{tabular}
\caption{Comparison of the cosmological parameter values (with $1\sigma$ errors) produced with a combination of CMB, supernovae data sets and DR8 quasars. In the third column we use all available $\ell$ bins in $30 \leq \ell \leq \ell_{\rm max}$, while in the fourth column we use only those bins in $10 \leq \ell \leq \ell_{\rm max}$ that satisfy a $1\sigma$-cut on ${\cal U}^{\alpha,\beta}_{\ell}$. We can see that quasars by themselves do not augment constraints on the cosmological parameters when we restrict ourselves to the $\Lambda$CDM model and use the WMAP9 and SN data sets. However, the change in $\Omega_m$, $h$, and $\sigma_8$ before and after we drop the systematics contaminated bins is interesting. This again shows the importance of removing bins that are contaminated by known and unknown systematics.}
\label{table:qsoresults}
\end{center}
\end{table*}

\begin{figure*}
\begin{center}
	\includegraphics[width=2.75in,angle=0]{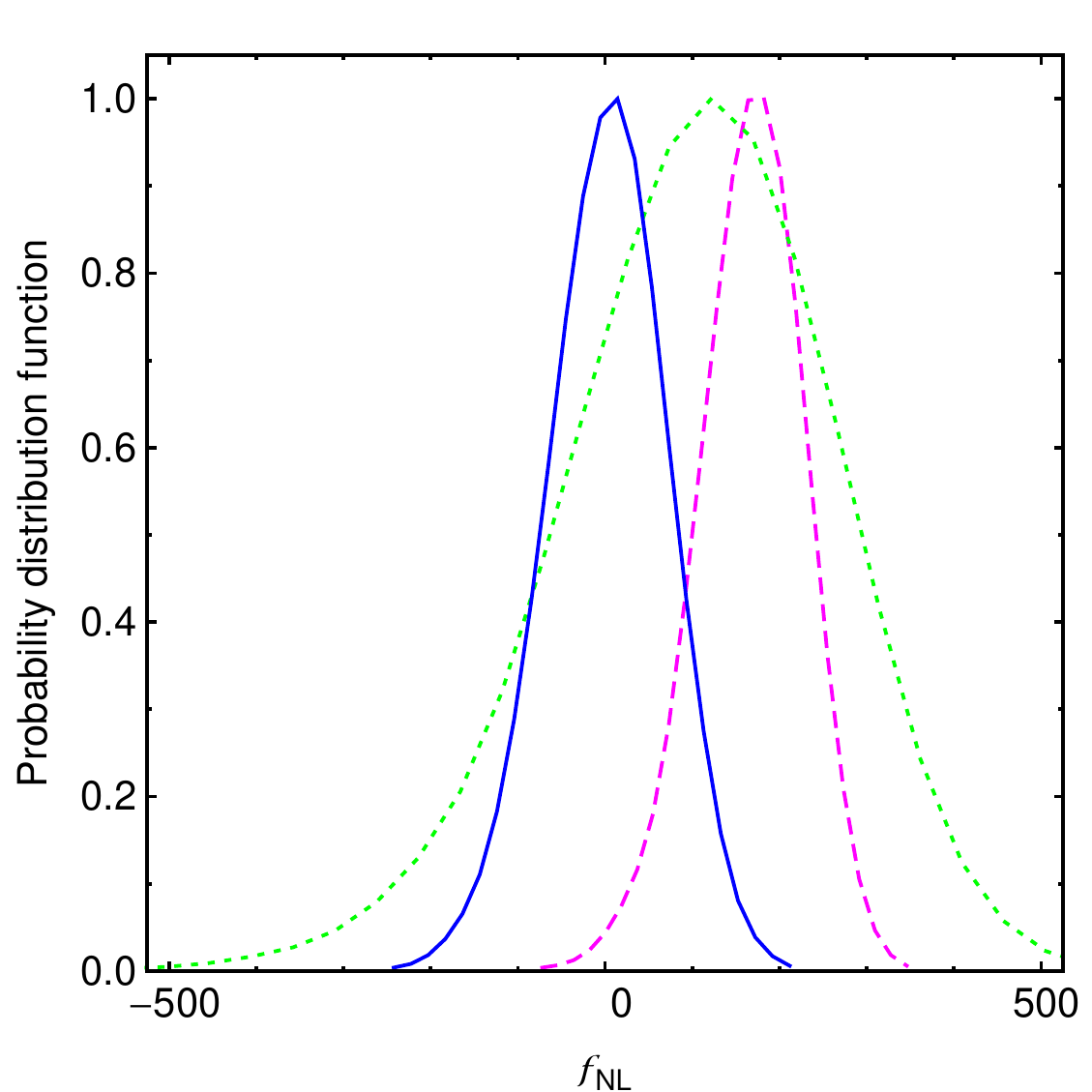}
	\caption{The marginalized probability distribution of $f_{\rm NL}$. The dashed magenta line shows the  marginalized distribution of $f_{\rm NL}$ when we combine WMAP9 + SN with QSO clustering over $\ell$ bins in the range $30 \leq \ell \leq \ell_{\rm max}$, using a ``typical'' choice to remove systematics in the angular power spectrum of quasars --- a hard cut of  $\ell \leq 30$. The dotted green line shows the marginalized distribution of $f_{\rm NL}$ when we instead remove contaminated bins as determined by cross-correlating different redshift  slices as described in \citep{agarwal13}. There is a decrease in constraining power when we remove more $\ell$ bins using the new method. The solid blue line combines WMAP9 + SN with QSOs and LRGs, again removing contaminated bins as determined by cross-correlating redshift slices \citep{agarwal13}. The quasar and LRG samples become consistent when we remove contaminated bins using our new method instead of the ``typical'' choice. For these results, we fixed the bias and $a$ parameters to their best-fit Gaussian values.}
\label{fig:comp_fnl}
\end{center}
\end{figure*}

\begin{figure*}
\begin{center}
	\includegraphics[width=2.75in,angle=0]{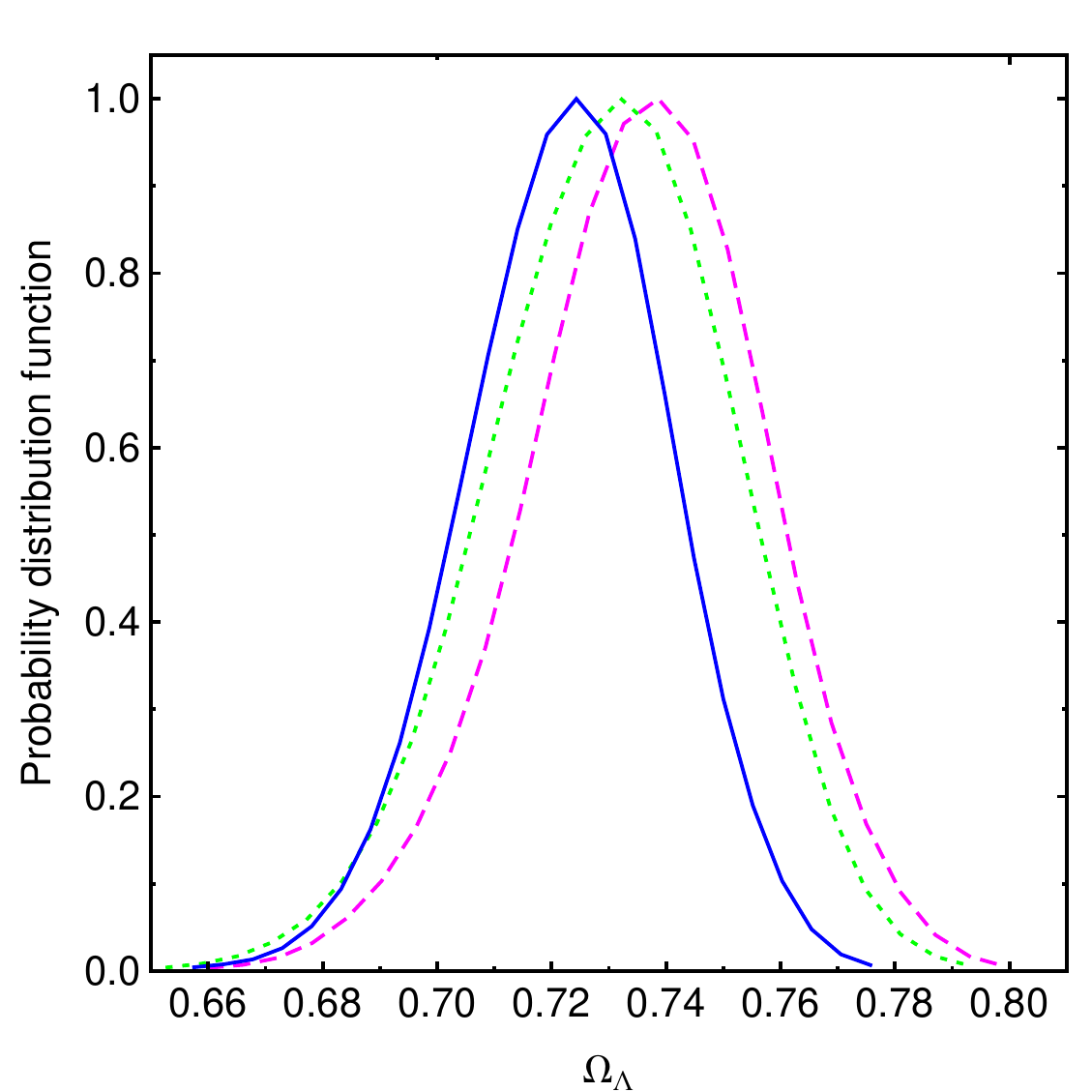}
	\caption{The marginalized probability distribution of $\Omega_\Lambda$. The color coding is similar to that in Fig.\ \ref{fig:comp_fnl}.}
\label{fig:comp_ol}
\end{center}
\end{figure*}

Next we present constraints on primordial non-Gaussianity using WMAP9 + SN + DR8 (QSO) data. In the presence of primordial (local) non-Gaussianity, the halo bias includes a scale-dependent term given by \cite{dalal08,matarrese08,Slosar:2008hx}
\bea
	\Delta b(M,k,z,f_{\rm NL}) & = & 3f_{\rm NL}[b_{1}(M,z) - p]\delta_{c} \frac{\Omega_{m}H_{0}^{2}}{k^{2}T(k)D(z)}.
\label{eq:deltabstd}
\eea
Here, $\delta_{c} \approx 1.686$ denotes the critical density for spherical collapse, $H_{0}$ is the Hubble constant, $T(k)$ is the matter transfer function normalized to unity as $k \rightarrow 0$, and $D(z)$ is the linear growth function normalized to $(1+z)^{-1}$ in the matter-dominated era. We add the above term to the Gaussian bias $b_{1}$ and introduce an extra parameter $f_{\rm NL}$ to the MCMC analysis. We set $p = 1.6$ for the extreme case that quasars only populate recently merged halos \cite{Slosar:2008hx} and use only those $\ell$ bins in $10 \leq \ell \leq \ell_{\rm max}$ that satisfy a $1\sigma$-cut on unknown systematics. We further include the DR8 (LRG) data of \cite{ross11, ho12}, with $p=1$ for LRGs, and use $\ell$ bins that satisfy a $3\sigma$-cut \cite{agarwal13} on unknown systematics. On marginalizing over all free parameters, including the bias (of quasars and LRGs) and shot noise-like $a$ parameters (for LRGs), we find that $f_{\rm NL} = -113^{+154}_{-154} \ (1\sigma {\rm \ error})$.

We next consider the case where the bias of quasars and galaxies is known, either from independent weak lensing or quasar/galaxy-CMB lensing cross-correlation measurements \citep{sherwin12}.  Given that the scales we use are relatively large, we can assume that effects of non-linear bias at these scales can still be relatively easy to be modeled with simulations.
 To test how well we can do in such a scenario where bias is obtained via a separate measurement, we constrain $f_{\rm NL}$ by fixing the bias and $a$ parameters to their best-fit Gaussian values, and performing an MCMC analysis over only the standard cosmological parameters and $f_{\rm NL}$. With quasar data alone, we find that using all $\ell$ bins in $30 \leq \ell \leq \ell_{\rm max}$ yields $f_{\rm NL} = 166^{+58}_{-57} \ (1\sigma {\rm \ error})$ while using only those bins in $10 \leq \ell \leq \ell_{\rm max}$ that satisfy a $1\sigma$-cut on unknown systematics produces $f_{\rm NL} = 103^{+148}_{-146} \ (1\sigma {\rm \ error})$. On further including the LRG sample we find that $f_{\rm NL} = 2^{+65}_{-66} \ (1\sigma {\rm \ error})$. For comparison, using only LRGs with a $3\sigma$-cut, and not including the quasar sample, yields $f_{\rm NL} = -17^{+68}_{-68} \ (1\sigma {\rm \ error})$. In Fig.\ \ref{fig:qsocltheory} we show the theoretical angular power spectra for quasars, using the best-fit and $68\%$ confidence values of $f_{\rm NL}$ obtained using quasars + LRGs, and fixing the bias and $a$ parameters to their best-fit values. The comparison of the marginalized distribution of $f_{\rm NL}$ and $\Omega_\Lambda$ between not dropping any of the contaminated bins, dropping those bins but including only quasars, and including LRGs are shown in Fig.\ \ref{fig:comp_fnl} and Fig.\ \ref{fig:comp_ol} respectively. It is apparent that once the contaminated bins are dropped, the quasar and LRG samples give consistent results. LRGs also tighten the constraints significantly. These results, along with the corresponding values for background cosmological parameters, are shown in Table \ref{table:qsofnlresults}.

\begin{table*}[!htb]
\begin{center}
\begin{tabular}{|c|c|c|c|}
\hline
Parameter & WMAP9 + SN & WMAP9 + SN & WMAP9 + SN \\
& + QSO & + QSO & + QSO + LRG \\
& (Before $1\sigma$-cut) & (After $1\sigma$-cut) & (After $1\sigma$- and $3\sigma$-cuts) \\
\hline
$f_{\rm NL}$ & $166^{+58}_{-57}$ & $103^{+148}_{-146}$ & $2^{+65}_{-66}$ \\
$\Omega_{b}h^{2}$ & $0.0227 \pm 0.0005$ & $0.0228 \pm 0.0005$ & $0.0227 \pm 0.0005$ \\
$\Omega_{\rm DM}h^{2}$ & $0.110 \pm 0.004$ & $0.112 \pm 0.004$ & $0.113 \pm 0.003$ \\
$\Omega_{m}$ & $0.264 \pm 0.020$ & $0.270 \pm 0.021$ & $0.277 \pm 0.017$ \\
$\Omega_{\Lambda}$ & $0.736 \pm 0.020$ & $0.730 \pm 0.021$ & $0.723 \pm 0.017$ \\
$h$ & $0.711 \pm 0.019$ & $0.707 \pm 0.019$ & $0.701 \pm 0.016$ \\
$\sigma_{8}$ & $0.801 \pm 0.021$ & $0.815 \pm 0.021$ & $0.823 \pm 0.014$ \\
$n_{s}$ & $0.973 \pm 0.012$ & $0.974 \pm 0.012$ & $0.972 \pm 0.012$ \\
\hline
\end{tabular}
\caption{Comparison of the cosmological parameter values (with $1\sigma$ errors) using CMB and supernovae data sets and using these data sets in combination with DR8 quasars and LRGs, in the presence of primordial (local) non-Gaussianity. In the second column we use all available $\ell$ bins in $30 \leq \ell \leq \ell_{\rm max}$, while in the third and fourth columns we use only those bins in $10 \leq \ell \leq \ell_{\rm max}$ that satisfy a $1\sigma$-cut (quasars) or $3\sigma$-cut (LRGs) on ${\cal U}^{\alpha,\beta}_{\ell}$. For these results, we fixed the bias and $a$ parameters to their best-fit Gaussian values.}
\label{table:qsofnlresults}
\end{center}
\end{table*}


\subsection{Companion Results}

In this paper DR8 quasars and LRGs are used to constrain primordial local non-Gaussianity. Our companion paper \cite{Agarwal:2013qta} studies constraints on a general form of the non-Gaussian halo bias, proportional to ${\cal A}_{\rm NL}/k^{\alpha}$, different from the usual $f_{\rm NL}/k^{2}$ form in Eq. (\ref{eq:deltabstd}) which holds only for the exact local ansatz. Since different models of inflation predict different forms of the scale dependence, this allows us to constrain models of inflation using a new observable, $\alpha$. We also present a Fisher matrix analysis using survey parameters consistent with DR8 to analyze the best constraints that can be obtained from a survey of this size. With current photometric large scale structure data,  the full marginalized upper limit on $\alpha$ is $2.0$ at the $95\%$ confidence level, consistent with the local ansatz.



\section{Conclusion and Discussion}
\label{sec:discuss}

We measured the angular clustering of quasars over the largest volume ever probed, coupled with the highest density of quasars ever used for such measurements. Although, in principle, extremely precise cosmological constraints can be obtained with a high volume and a high density of tracers, this study demonstrated the necessity and importance of detecting and removing unknown systematics.  

In the case of photometric quasars, even after removing a large number of systematics using cross-correlations between systematics maps and quasar density maps, significant contamination remains. We applied a new method \cite{agarwal13} to detect the magnitude with which unknown systematics affect quasar density maps. This method uses cross-correlations between different redshift slices of quasars, relying on the fact that overlapping quasars from different redshift bins produce a relatively modest signal that can be adequately modeled within the current cosmological framework. Once we determined the magnitude of the contamination, we removed those ``bins'' ($\ell$ bins) which are contaminated. 

Our method is not perfect, since there is insufficient information to disentangle which redshift slice in a cross-correlation is the main cause of any systematic contamination. Therefore, we removed $\ell$ bins that are contaminated from both redshift slices, even though maybe only one of the redshift slices is truly contaminated. Some other shortcomings of the method presented in \cite{agarwal13} include errors in the ``true'' cross-power arising from uncertainty in the quasar redshift distribution and not accounting for errors in the observed cross-power, both of which will lead to an error in our estimate of ${\cal U}_{\ell}^{\alpha,\beta}$. Nonetheless, this approach provides a mechanism with which to estimate conservatively which $\ell$ bins are contaminant-free. Further, this method can be used at any chosen tolerance to unknown systematics --- convenient as different users of our method may have different science goals.

We determined conservative constraints on local primordial non-Gaussianity from the angular power-spectrum of both quasars and LRGs \citep{ross11, ho12}, finding that $f_{\rm NL} = -113^{+154}_{-154}$ $(1\sigma {\rm \ error})$, on discarding bins from both samples that were contaminated with unknown systematics, and marginalizing over all background cosmology parameters including the bias (of quasars and LRGs) and shot noise-like $a$ parameters (for LRGs). We then checked how our constraints changed on fixing the bias and $a$ parameters to their best-fit Gaussian values. With quasar and LRG data combined, and still removing contaminated bins, we found a constraint of $f_{\rm NL} = 2^{+65}_{-66} \ (1\sigma {\rm \ error})$. This constraint is  comparable to \cite{Slosar:2008hx}, whose constraints were derived primarily from a single redshift slice of quasars which was deemed to be relatively systematics-free. Our results are also consistent with (though significantly weaker than) other recent publications on primordial non-Gaussianity from large scale structure and the CMB \citep{Ade:2013ydc,Giannantonio:2013uqa,Leistedt:2014zqa}. We conclude that in order to best use future large scale structure data, it is important to develop robust techniques to handle both known and unknown systematics, and additionally methods to independently probe the bias of dark matter halos.


\section{Acknowledgments}


S. H. would like to thank Eiichiro Komatsu, Tommaso Giannantonio, and Robert Lupton for many discussions concerning both the theory and the data. S. H. and N. A. also thank Sarah Shandera for many useful discussions. 

S. H. is partially supported by the New Frontiers in Astronomy and Cosmology program at the John Templeton Foundation and was partially supported by RESCEU fellowship, and the Seaborg and Chamberlain Fellowship (via Lawrence Berkeley National Laboratory) during the preparation of this manuscript. N. A. is supported by the McWilliams fellowship of the Bruce and Astrid McWilliams Center for Cosmology. A. D. M. is a research fellow of the Alexander von Humboldt Foundation of Germany and was partially supported through NSF Grant 1211112 and NASA ADAP award NNX12AE38G. N. A. and R. O. are both partially supported by the New Frontiers in Astronomy and Cosmology program at the John Templeton Foundation.

Funding for SDSS-III has been provided by the Alfred P. Sloan Foundation, the Participating Institutions, the National Science Foundation, and the U.S. Department of Energy Office of Science. The SDSS-III web site is http://www.sdss3.org/. SDSS-III is managed by the Astrophysical Research Consortium for the Participating Institutions of the SDSS-III Collaboration including the University of Arizona, the Brazilian Participation Group, Brookhaven National Laboratory, University of Cambridge, Carnegie Mellon University, University of Florida, the French Participation Group, the German Participation Group, the Instituto de Astrofisica de Canarias, the Michigan State/Notre Dame/JINA Participation Group, Johns Hopkins University, Lawrence Berkeley National Laboratory, Max Planck Institute for Astrophysics, New Mexico State University, New York University, Ohio State University, Pennsylvania State University, University of Portsmouth, Princeton University, the Spanish Participation Group, University of Tokyo, University of Utah, Vanderbilt University, University of Virginia, University of Washington, and Yale University.


\bibliography{references}
\bibliographystyle{JHEP}


\end{document}